\documentclass[useAMS,usenatbib]{mn2e}

\usepackage{amsmath} 
\usepackage{amssymb} 
\usepackage{graphicx}

\newcommand{\lya}{{Lyman-$\alpha$~}}
\newcommand\ion[2]{#1$\;${\scshape{#2}}}
\newcommand{\Msol}{{$M_{\odot}$}}
\newcommand{\fixbib}[1]{}

\title[$z \ga 6$ QSO near-zones]{Probing the end of reionization with  the near-zones of $z \ga 6$ QSOs}

\author[L.C. Keating et al.]{\parbox{\textwidth}{Laura C. Keating$^{1,2}$\thanks{E-mail:
    lck35@ast.cam.ac.uk}, Martin G. Haehnelt$^{1,2}$, Sebastiano Cantalupo$^{3}$ \& Ewald Puchwein$^{1,2}$\\ }\vspace{0.4cm} \\
\parbox{\textwidth}{$^1$Institute of Astronomy, University of Cambridge,
  Madingley Road, Cambridge, CB3 0HA, UK\\
$^2$Kavli Institute  for Cosmology,  University of Cambridge,
  Madingley Road, Cambridge, CB3 0HA, UK\\
$^3$Institute for Astronomy, Department of Physics, ETH Zurich,
   Wolfgang-Pauli-Strasse 27, 8093, Zurich, Switzerland}
}

\begin{document}

\date{Accepted 2015 August 28. Received 2015 August 11; in original form 2015 June 10\\
This is a pre-copyedited, author-produced version of an article accepted for publication in MNRAS following peer review.}

\pagerange{\pageref{firstpage}--\pageref{lastpage}} 

\pubyear{2014}

\maketitle

\label{firstpage}

\begin{abstract}
QSO near-zones are an important probe of the the ionization state of the IGM at $z\sim 6-7$, at the end of reionization. We present here high-resolution cosmological 3D radiative transfer simulations of  QSO environments for a wide range of host halo masses, $10^{10-12.5}$\Msol.  Our  simulated near-zones reproduce both the overall decrease of observed near-zone sizes at $6 <z < 7 $ and their scatter. The observable near-zone properties in our simulations depend only very weakly  on the mass of the host halo.  The size of the \ion{H}{ii} region expanding into the IGM is generally limited by \mbox{(super-)}Lyman Limit systems loosely associated with  (low-mass) dark matter haloes. This leads to a strong dependence of near-zone size on direction and drives the large observed scatter. In the simulation centred on our most massive host halo, many sightlines  show strong red damping wings even for   initial volume averaged neutral hydrogen fractions as low as $\sim  10^{-3}$. For QSO lifetimes long enough to allow growth of the central supermassive black hole while optically bright, we can reproduce the observed near-zone of ULAS J1120+0641  only with an IGM that is initially  neutral. Our results suggest that larger samples  of  $z>7$ QSOs will provide important constraints on the evolution of the neutral hydrogen fraction and thus on how late reionization ends. 
\end{abstract}

\begin{keywords}
galaxies: high-redshift -- quasars: absorption lines -- intergalactic medium -- dark ages, reionization, first stars
\end{keywords}

\section{Introduction}

An important unsolved problem in modern cosmology is understanding how our Universe transitioned from the ``dark ages'', following recombination, to the ionized Universe we observe today. Despite much progress in the last decade there is still  considerable uncertainty with regard to the exact timing and  topology of reionization and which sources are the main contributors to the ionizing emissivity. The evolution of the Lyman-$\alpha$ (Ly$\alpha$) forest and detection of a Gunn-Peterson trough blueward of Ly$\alpha$ suggest that the neutral fraction of the intergalactic medium (IGM) is rapidly increasing for $z>5.7$ \citep{fan2006,becker2015}. Apparent drops in the number of \lya\ emitters at $z = 6-7$  also suggest a rapidly increasing neutral fraction of hydrogen \citep[e.g.,][]{stark2010,ouchi2010,finkelstein2013,pentericci2014}. Interestingly, the latest measurements of the optical depth due to Thomson scattering from observations of the cosmic microwave background (CMB) are significantly  lower  than earlier measurements \citep{planck2013,planck2015}.   For the rather artificial assumption of instantaneous reionization, the new inferred reionization redshift from the latest CMB measurements is $z_{\textnormal{\scriptsize{reion}}} = 8.8^{+1.2}_{-1.1}$. For more realistic extended evolutionary histories of the neutral hydrogen  fraction, the new measurements suggest that reionization  ends  somewhat  later than previously thought \citep{haardtmadau2012,choudhury2014,robertson2015,bauer2015,chardin2015}. 

With  a number of near-infrared surveys under way to further push the redshift barrier beyond $z \gtrsim 7$, this makes QSO near-zones a promising tool to further investigate how reionization ended. Near-zones are defined as regions of transmitted flux blueward of the Ly$\alpha$ emission, extending from the source to the point where the flux drops below 10 per cent for the first time \citep{fan2006}. UKIDSS, VISTA and Pan-STARRS are now returning the first discoveries of $z \gtrsim 6.5$ quasars \citep{mortlock2011,venemans2013,venemans2015}. The near-zones of these bright QSOs are a unique observational tool to study the high-redshift IGM, as they allow us to push past the redshift at which the opacity in the Ly$\alpha$ forest becomes too large to extract detailed information \citep[e.g.,][]{fan2006,carilli2010}. 

\citet{carilli2010} found evidence for decreasing near-zone size with increasing redshift in the range $5.7 < z <6.4$, suggesting evolution in the neutral fraction of the IGM, as well as substantial scatter in the observed near-zone sizes. This trend continues to the current highest redshift QSO, ULAS J1120+0641, at $z=7.085$ (see \citet{venemans2015,wu2015} and \citet{reed2015} for more recent updates with additional QSOs  at $z>6$). 

The small near-zone size of ULAS J1120+0641, together with the possible discovery of a  damping wing observed redward of Ly$\alpha$ transmission, also suggests that the IGM at $z\sim 7$ may be significantly more neutral than at $z \sim 6$ \citep{mortlock2011,bolton2011}, but see \citet{bosman2015} for a reassessment of the significance of the evidence for suppressed  \lya emission in this QSO by a significantly neutral surrounding IGM.   

The relation of the sizes of near-zones to the neutral fraction of the surrounding IGM is, however, not straight forward. The size of the near-zone generally underestimates the size of the \ion{H}{ii} region \citep[e.g.,][]{bolton2007a,maselli2007}. \citet{maselli2009} showed that in a sample of $z \sim 6$ quasars, all the observed near-zones sizes appear to be due to the proximity zone around the quasar, rather than tracing the \ion{H}{ii} region.  Reionization is of course  not expected to be a spatially homogeneous process \citep{mhr2000,furlanetto2005} and, as quasars may form in biased, high-density regions, the surrounding IGM may already be substantially ionized by the time the quasar turns on \citep{lidz2007}. 

It is also not clear to what extent the environment and, in particular, the mass of the host halo influence the observed properties of high-redshift QSO near-zones \citep{lidz2007,wyithe2008}. Unfortunately observational constraints on the environment and host halo masses of high-redshift QSOs  so far have remained inconclusive. \citet{banados2013} and \citet{simpson2014} find no evidence for the overdensities that would be expected around the most massive haloes, while \citet{morselli2014} claim to observe an overdense environment around four $z \sim 6$ QSOs. The mass of the black hole powering ULAS J1120+0641 is estimated to be $(2.0 \pm ^{1.5}_{0.7}) \times 10^{9} \, M_{\odot}$ \citep{mortlock2011}. Based on  high-resolution numerical simulations of the growth of galaxies and black holes in a  cosmological context, it has been shown that it is possible to form  a black hole of this mass by growth limited to the Eddington accretion rate from a seed of $10^{5} \, M_{\odot}$ at $z \sim 15$  but these black holes will only be found in the most massive dark matter haloes (with $M_{\textnormal{\scriptsize{halo}}} \sim 10^{12} \, M_{\odot}$) at $z \sim 6$ \citep{sijacki2009,costa2014}. Conversely, \citet{fanidakis2013} use semi-analytic modelling allowing for super-Eddington accretion rates to suggest that luminous quasars are  found in less massive haloes; however, their modelling fails to reproduce the bright end of the high-redshift quasar luminosity function, as $10^{9} \, M_{\odot}$ black holes are not formed until $z < 4$. We  therefore simulate here three different regions of the IGM at $z=7$, centred on host haloes spanning a wide range of masses from $10^{10}$ to $10^{12}$ \Msol\ and covering a range of environments from an average density region to a very highly biased proto-cluster region. 

This work follows on  from the work of \citet{bolton2011} but differs in two important respects: first, we test a range of host halo masses and environments extending  to a rather large host halo mass and highly biased environments predicted by realistic models for the growth of billion solar mass black holes believed to power bright high-redshift QSOs; second, we use the  3D radiative transfer code \textsc{radamesh} \citep{cantalupo2011} which allows us to sample the near-zones effectively in all directions.  By choosing different background photo-ionization rates, we vary the initial neutral fraction of the gas in the surrounding IGM before the QSO turns on.

This paper is organised as follows. In section 2, we describe the numerical simulations we used to model the ionization fronts around high-redshift QSOs. In section 3, we investigate the effect of different environments and initial neutral fractions on the QSO near-zones. In section 4, we compare our simulations to the observed near-zones at $z=6-7$ and consider how best to constrain the neutral fraction at $z \gtrsim 6.5$. Finally, in section 5, we present  our conclusions. We use the cosmological parameters $h = 0.73$, $\Omega_m=0.25$, $\Omega_{\Lambda}=0.75$  and $\Omega_{b}h^2=0.024$. The prefixes `p' and `c' before distance measures refer to proper and comoving distances respectively.

\section{Simulations of QSO Near-Zones}

\subsection{Hydrodynamical Simulations of High-Redshift Haloes}

\begin{figure*}
\includegraphics[trim={0cm 0cm 1.5cm 7cm},clip, width=2.1\columnwidth]{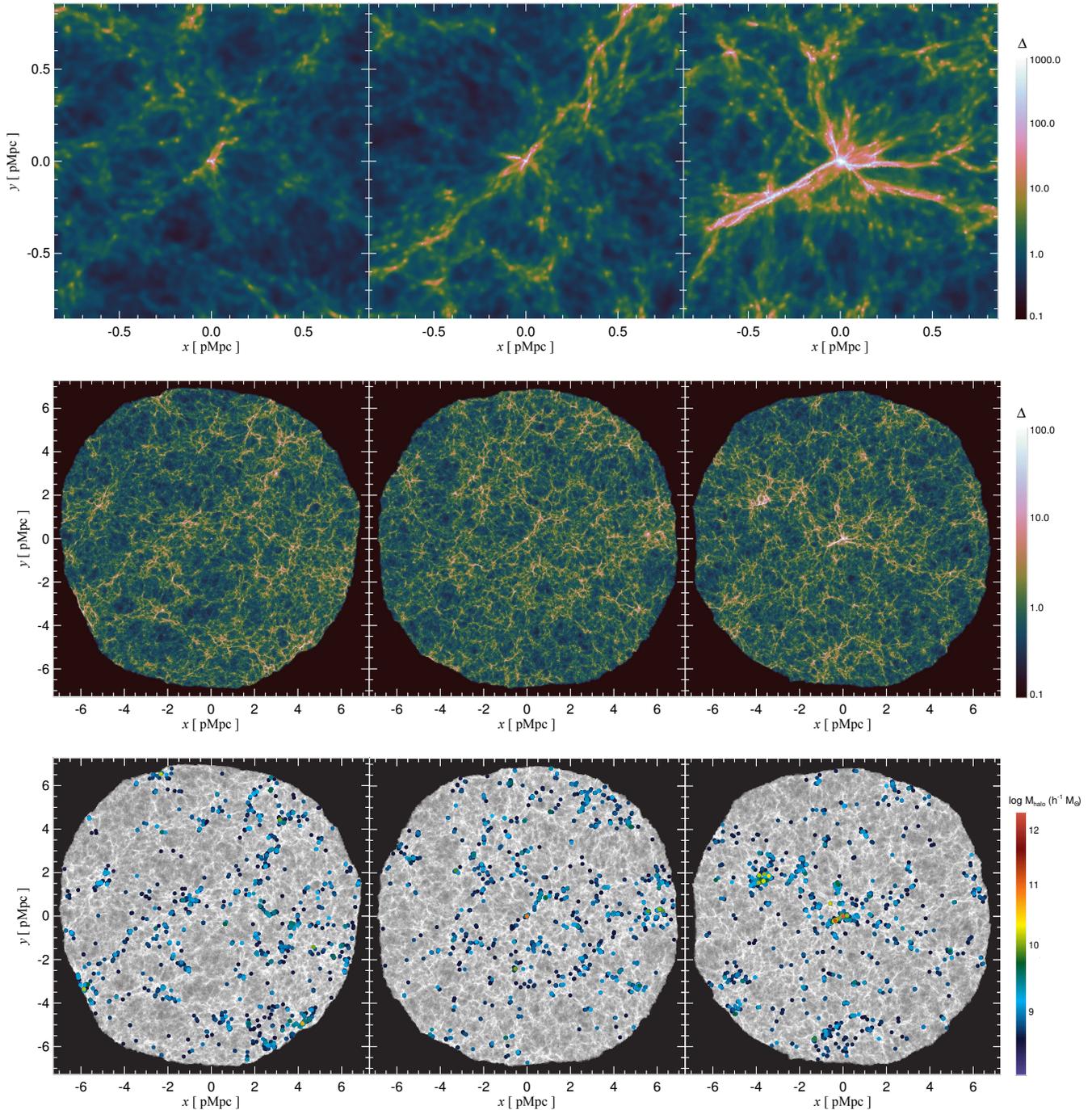}
\caption{Maps of the projected overdensity for the three simulated regions at $z=7$: average (left), intermediate (middle) and overdense (right). The average region contains some thin filaments as well as large voids, while the intermediate region contains more pronounced filaments and also smaller voids. The overdense region contains several prominent filaments surrounding the central massive halo. The top panel shows the 10 $h^{-1}$ cMpc surrounding the host halo, the middle panel shows a slice through the entire region and the bottom panel shows the projected overdensity with the dark matter haloes contained in a slice of the same thickness overlaid.}
\label{densities}
\end{figure*}

\begin{figure*}
\includegraphics[trim={1.5cm 12.5cm 1.5cm 9.5cm},clip, width=2.1\columnwidth]{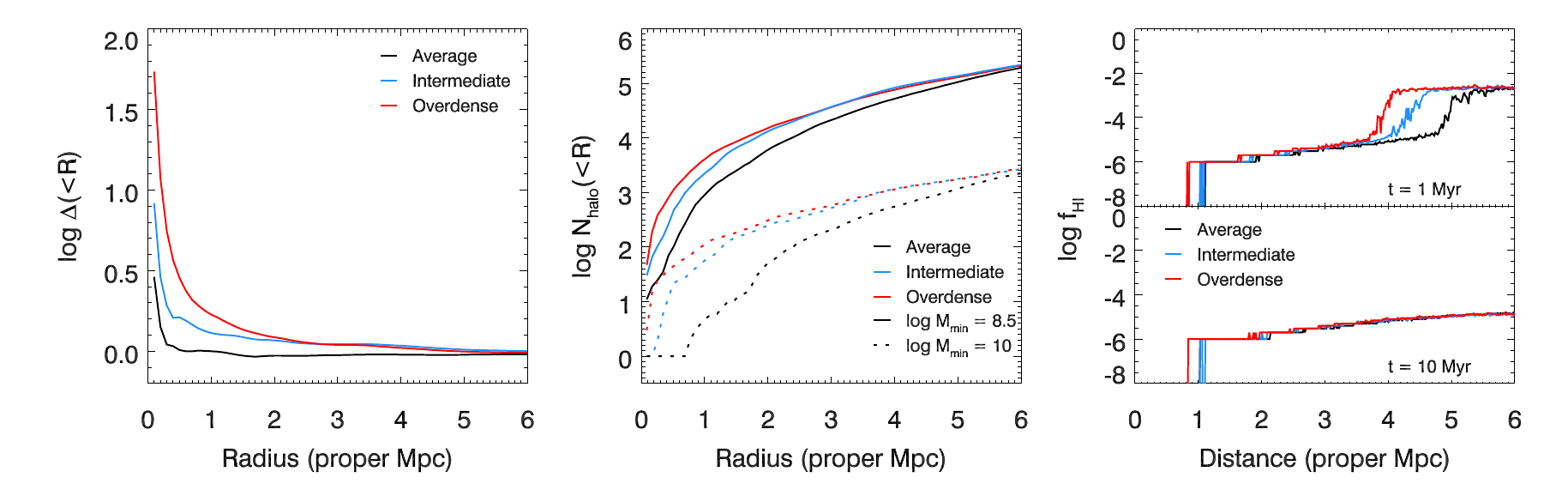}
\caption{Left: Overdensity within a sphere with radius specified on the horizontal axis, centred on the host halo, for the three regions. Outside the central 2 pMpc, the intermediate and overdense regions show similar densities that are still slightly above the average cosmic density, while the density of the average region lies much closer to the cosmic density at $z=7$. Beyond a radius of 5  pMpc, all three regions display similar overdensities. Middle: The number of haloes above a cut-off mass $\log M_{\textnormal{\scriptsize{Halo}}} (h^{-1} M_{\odot})  =$ 8.5 (solid line) and 10 (dotted line) contained in a sphere with radius specified on the horizontal axis for the three simulations. Right: The median neutral fraction over all sightlines for our three regions with initial volume-weighted average neutral fraction  $\langle f_{\textnormal{\scriptsize{\ion{H}{i}}}} \rangle_{\scriptsize{\textnormal{init}}} = 0.02$ after 1 Myr (top panel) and 10 Myr (bottom panel).}
  \label{haloes}
\end{figure*}

As the mass of the dark matter haloes that bright QSOs lie in at $z \sim 6- 7$ is still  uncertain, we have simulated a range of cosmological environments, following the approach in \citet{costa2014}: an ``average'' region, centred on a $10^{10} h^{-1} M_{\odot}$ halo, an ``intermediate'' region, centred on  a $1.7\times 10^{11} h^{-1} M_{\odot}$ halo and an ``overdense'' region, centred on  a $2.5 \times 10^{12} h^{-1} M_{\odot}$ halo. To model these three distinct regions, we resimulated three haloes taken from a snapshot of the Millennium simulation \citep{Springel2005mill} at $z \sim 6$, using the TreePM-SPH code \textsc{p-gadget3} (last described in \citealt{Springel2005gadget}). The Millennium simulation is a dark-matter only simulation of a cube with side length $500 h^{-1}\, \textnormal{Mpc}$ run from redshift 127 up to the present time. The large volume of the simulation allows for formation of rare objects, such as the most massive halo we simulate. We used the group catalogues to select haloes of desired mass at $z=6$ and traced all particles in a radius around these haloes back to $z=127$ to generate our initial conditions. We included a factor of $4^3$ times more particles than the parent simulation in a high-resolution region, which was also populated with gas particles. Outside this region, increasingly lower resolution dark matter particles were also simulated so as to correctly capture the large-scale tidal forces while minimising computational cost. The spatial resolution of our simulation is a factor 3 lower than previous studies of $z \sim 7$ near-zones by \citet{bolton2011}. Unfortunately, this  is an unavoidable compromise that had to made in order to run large boxes containing the rare, massive haloes in which we are interested.
\begin{table}
\centering
\begin{tabular}{c|c|c|c}
Region&$M_{\textnormal{\scriptsize{Halo}}} (\times 10^{12} \, h^{-1} \, M_{\odot})$&$\Delta_{\textnormal{\scriptsize{r=0.5}}}$&$\Delta_{\textnormal{\scriptsize{r=3}}}$\\
\hline
Average&0.01&1.00&0.95\\
Intermediate&0.17&1.63&1.11\\
Overdense&2.50&2.86&1.10\\
\end{tabular}
\caption{Here we list the mass of the host halo and the overdensity $\Delta = \rho / \bar{\rho}$ (where $\bar{\rho}$ is the critical density) of a sphere with radius 0.5 and 3 pMpc, centred on the host halo, for the three regions we simulate at $z=7$.}
\label{regions}
\end{table}

We model star formation in the same way as \citet{viel2004}, such that gas particles with density greater than 1000 times the mean baryon density and a temperature less than $10^{5}$ K are turned into collisionless star particles.  Simulations with this prescription have been shown to reproduce well the low-density IGM responsible for the bulk of the absorption in the high-$z$ \lya\ forest, while providing increased numerical efficiency. We have also investigated a simulation with a more  physically motivated star formation prescription that also included the effect of stellar and AGN feedback, which we discuss in Appendix \ref{sec:feedback}.

The radius of our high-resolution region is approximately 40 $h^{-1}$ comoving Mpc (the exact radius of each simulation depends on the virial radius of the halo of interest) and the mass of the gas particles is $2.8 \times 10^6 \, h^{-1} \,\textnormal{M}_{\odot}$. The softening length is 1.25 $h^{-1}$ comoving kpc. The haloes contain at least 32 dark matter particles and the virial temperature of the smallest haloes is $T_{\textnormal{\scriptsize{V}}} \sim 650$ K for $\mu = 0.61$. 

The \citet{haardtmadau2012} model of the UV background was used, assuming the optically thin limit for ionizing radiation. Note that although we will later recompute the ionization state of the gas for a series of different photo-ionization rates, the change  of UV background should in principle affect the temperature of the gas, an effect which we neglect when we reprocess the simulations.  

Examples of the density field in the immediate vicinity of the three host haloes and in a slice through the full simulation volume are shown in the top and middle panels of Figure \ref{densities}. The average density region is characterised by thinner  filaments of gas and large voids, while the overdense region shows more prominent filaments and higher density diffuse gas. Within the central 2 pMpc, the difference in the mean overdensity of the three regions is evident (see also the left panel of Figure \ref{haloes}), with the mean overdensity of the region increasing  with increasing mass of the host halo as  expected. Beyond this distance, the intermediate and overdense regions display similar mean densities, while the mean density of the average region remains lower until a radius of 5 pMpc. This is described more quantitatively in Table \ref{regions}. 

In the bottom panel of Figure \ref{densities}, we plot the distribution of the subhaloes in a slice with thickness 0.03 pMpc. Note that the most massive  halo in the average region is not our host halo (see, for example, the $8 \times 10^{10} \, h^{-1} \, M_{\odot}$ halo in the lower right of the bottom left image in Figure  \ref{densities}). However, in the region surrounding the host halo, the distribution of haloes is clearly different (middle panel of Figure \ref{haloes}). We find that the regions centred on  more massive haloes also contain more haloes of smaller masses. This is true for both haloes with masses above $\log M_{\textnormal{Halo}} (h^{-1} M_{\odot}) = 8.5$ and 10. Similar to the trend with overdensity in the left panel of Figure \ref{haloes}, we find that the number of haloes in the region is increasing with the mass of the host halo up to a radius of about 2 pMpc, after which the intermediate and overdense regions become very alike.

\subsection{Modelling QSO Ionization Fronts}
\label{sec:model_qso}

To calculate the neutral fraction of the gas after it is ionized by the QSO, we post-processed the simulations with the 3D radiative transfer code \textsc{radamesh}. {\sc radamesh} (Radiative transfer on an ADAptive MESH) is a three-dimensional radiative transfer code, described in detail in \citet{cantalupo2011}. It makes use of a photon-conserving, ray-tracing approach that is adaptive in space and time. This grid-based algorithm combines elements from two common techniques for solving radiative transfer problems: the Monte Carlo approach and the method of characteristics. This scheme allows for accurate and computationally efficient solutions of the radiative transfer equation. By tracking the path of ionizing radiation over a static density field, the time evolution of the neutral hydrogen and helium fractions, as well as the temperature, can be calculated. It has been tested on the standard tests for radiative transfer codes from \citet{iliev2006} and shown to perform well.

Fully coupling simulations of radiative transfer to the hydrodynamical simulations is a computationally expensive problem, but fortunately for the problem considered here, this is of little relevance. For the expansion of a \ion{H}{ii} region around a luminous QSO, the relevant time scale for the motion of the ionization front is much shorter than the hydrodynamical time scales for the gas. We have also investigated the timescale for photoevaporation of gas in haloes with masses equivalent to the minimum mass our simulation resolves using the method outlined in \citet{iliev2005}, and find that the time for evaporation is on the order of 100 Myr for a halo with $M_{\textnormal{\scriptsize{Halo}}} = 10^8 h^{-1} M_{\odot}$ at a radius 10 comoving Mpc from the source. As the time interval over which we evolve the radiative transfer is much shorter than this, we do not expect photoevaporation to be an issue.

For simplicity and for computational speed, we model the initial ionization state of the IGM using a homogeneous UV background together with a self-shielding model to account for optically thick absorbers.  The effect of self-shielding on the initial ionization state was implemented  in post-processing using a simple self-shielding density threshold implemented similarly to \citet{rahmati2013}. Although fluctuations in the UV background are expected to be smaller at low redshift, when the mean free path of an ionizing photon is short, as expected for $z>6$ \citep[e.g.,][]{songaila2010}, these fluctuations can become large and modelling observations with a uniform UV background can become difficult \citep{becker2015,chardin2015}. It would obviously be more realistic when considering the ionization state of the IGM to take into account the contribution to the ionizing background of individual sources, however to model this correctly would add an additional level of complexity to this problem and is beyond the scope of this paper.

To perform the radiative transfer, outputs from the hydrodynamical simulations at $z=6$ and 7 were divided into octants and mapped onto multi-mesh grids, each with base grid size $512^3$. The grid was refined wherever 16 or more gas particles were present in a cell, resulting in four levels of refinement. Cells on the highest level of refinement had a spatial resolution of 0.95 pkpc. The initial temperature of the gas was taken from the hydrodynamical simulation. Collisional ionizations are also taken into account. Taking the same values for the properties of ULAS J1120+0641 as \citet{bolton2011}, we placed a source with ionizing photon rate $\dot{N} = 1.3 \times 10^{57} \, \textnormal{s}^{-1}$ and spectral index $\alpha = 1.5$ in the densest cell of the central halo. By keeping the properties of the QSO fixed and varying the mass of the host halo and the redshift, we were able to estimate the effect that the mass of the host halo and the evolution of the gas distribution has on the extent of the near-zones.

We consider a range of initial ionization states for the gas, giving volume-weighted initial neutral fractions ranging from $\langle f_{\textnormal{\scriptsize{\ion{H}{i}}}} \rangle_{\scriptsize{\textnormal{init}}} \sim 10^{-4}$ to $1$ for the neutral fraction in the vicinity of the quasar. The photo-ionization rates and corresponding volume-weighted neutral fractions are summarised in Table \ref{photoi}. Photo-ionization rates for \ion{He}{i} and \ion{He}{ii} were calculated by rescaling to match the ratios between the photo-ionization rates in \citet{haardtmadau2012}.

\begin{table}
\centering
\begin{tabular}{c|c|c}
$\log \Gamma_{\textnormal{\scriptsize{\ion{H}{i}}}}$ (s$^{-1}$)&$\langle f_{\textnormal{\scriptsize{\ion{H}{i}}}}(z=6) \rangle_{\textnormal{\scriptsize{init}}}$&$\langle f_{\textnormal{\scriptsize{\ion{H}{i}}}}(z=7) \rangle_{\textnormal{\scriptsize{init}}}$\\
\hline
$-17.0$&$9.78 \times 10^{-1}$&$9.82 \times 10^{-1}$\\
$-14.8$&$5.21 \times 10^{-2}$&$1.11 \times 10^{-1}$\\
$-14.3$&$9.42 \times 10^{-3}$&$1.99 \times 10^{-2}$\\
$-13.7$&$1.43 \times 10^{-3}$&$2.20 \times 10^{-3}$\\
$-12.8$&$1.14 \times 10^{-4}$&$1.68 \times 10^{-4}$\\
\end{tabular}
\caption{\ion{H}{i} photo-ionization rates used to calculate the initial ionization state of the gas with the \citet{rahmati2013} model for self-shielded absorbers at $z=6$ and 7. The volume weighted neutral hydrogen fraction as calculated in the overdense region is also shown.}
\label{photoi}
\end{table}

Ionizing UV radiation was sampled from twenty logarithmically spaced energy bins, ranging from 1 to 8 Ryd. A time-dependent, non-equilibrium chemistry solver computes the number density of six species (\ion{H}{i}, \ion{H}{ii}, \ion{He}{i}, \ion{He}{ii}, \ion{He}{iii}, $e^{-}$) as well as the temperature of the gas. Recombination radiation was treated using an ``on-the-spot'' approximation, whereby ionizing photons are absorbed in the same cell in which they were emitted (i.e., that the mean free path of the ionizing photon is smaller than the minimum length scale resolved by our simulation) and that the photons are absorbed by the same species from which they were emitted.  

We further assume  the speed of light in our simulation to be  infinite. Although this means that the speed that the \ion{H}{ii} region initially expands at can appear to exceed the speed of light, it can be shown that it is a good description of how the front propagates from the point of view of an observer for spectra along the line of sight \citep{white2003,yu2005,shapiro2006,bolton2007a,cantalupo2008}. The QSO lifetime $t_{\textnormal{\scriptsize{Q}}}$ that we will refer to throughout this paper is the time at which the photons were emitted from the source. This can be related to the time $t'$ by $t_{\textnormal{\scriptsize{Q}}} = t' - R(t')/c$, where $R(t')$ is the distance that has been travelled in a time $t'$ and $c$ is the speed of light. 

\section{QSO Near-Zone Sizes}

\subsection{Evolution of the \ion{H}{II} Regions}
\label{sec:fronts}

\begin{figure*}
\includegraphics[trim={0cm 0cm 1cm 1.5cm},clip,width=1.9\columnwidth]{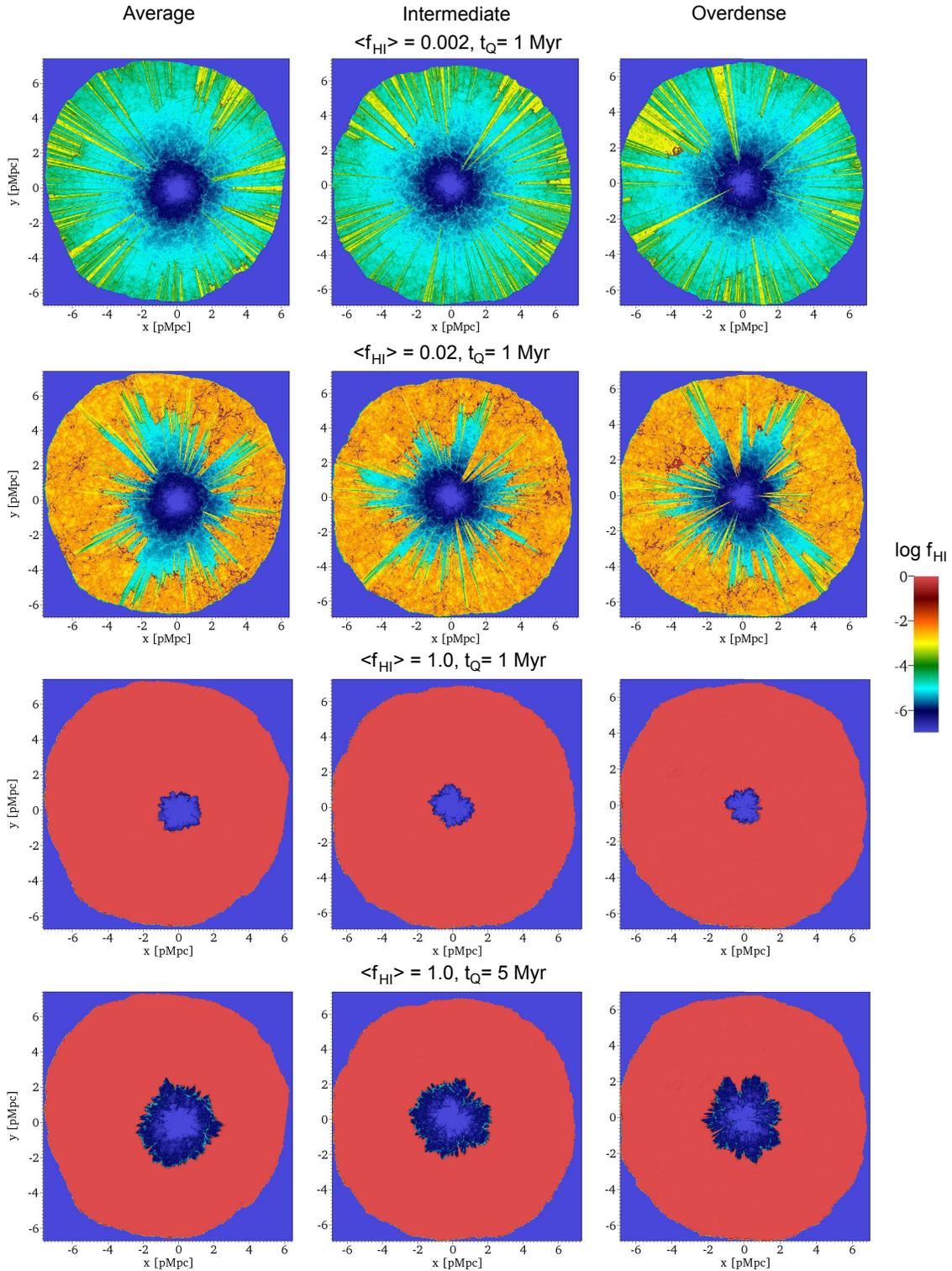}
\caption{Maps of the projected neutral fraction for average (left), intermediate (middle) and overdense (right) regions over a slice 0.03 pMpc thick at $z=7$. The top row shows simulations after $t_{\textnormal{\scriptsize{Q}}} = 1$ Myr with initial volume-weighted neutral fraction $\langle f_{\textnormal{\scriptsize{\ion{H}{i}}}} \rangle_{\textnormal{\scriptsize{init}}} = 0.002$, second row has $\langle f_{\textnormal{\scriptsize{\ion{H}{i}}}} \rangle_{\textnormal{\scriptsize{init}}} = 0.02$ and third row has $\langle f_{\textnormal{\scriptsize{\ion{H}{i}}}} \rangle_{\textnormal{\scriptsize{init}}} = 1.0$. The fourth row has  $\langle f_{\textnormal{\scriptsize{\ion{H}{i}}}} \rangle_{\textnormal{\scriptsize{init}}} = 1.0$ and is shown at  $t_{\textnormal{\scriptsize{Q}}} = 5$ Myr. Note that this is not an instantaneous picture, but rather shows the size of the \ion{H}{ii} region that would be measured by an observer along each sightline at fixed quasar lifetime.}
 \label{fhi_regions}
\end{figure*}

\begin{figure*}
\includegraphics[trim={0cm 0cm 1.5cm 21cm},clip,width=2.1\columnwidth]{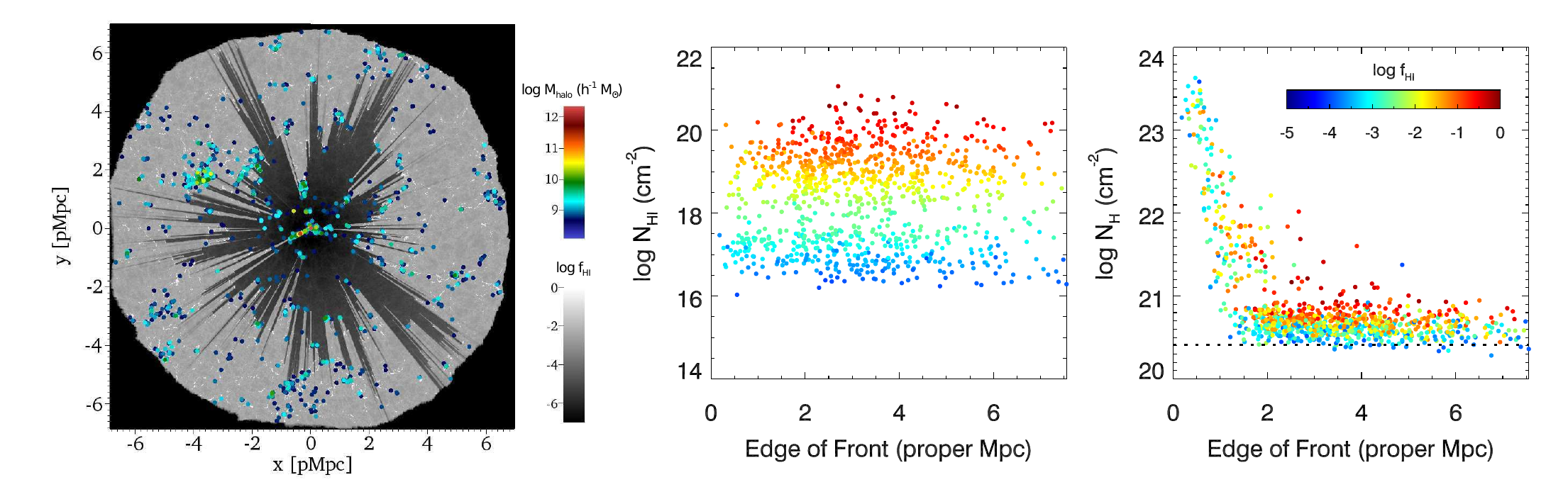}
\caption{Left: The projected neutral fraction over a slice with thickness 0.03 pMpc for the overdense region with initial neutral fraction $\langle f_{\textnormal{\scriptsize{\ion{H}{i}}}} \rangle_{\textnormal{\scriptsize{init}}} = 0.02$. (Sub)haloes in a slice of the same thickness found using \textsc{subfind} are overplotted. Middle:  \ion{H}{i} column density at the position where the near-zone is found over a range  $\pm 500$ km s$^{-1}$  after $t_{\textnormal{\scriptsize{Q}}} = 1$ Myr in our overdense region with $\langle f_{\textnormal{\scriptsize{\ion{H}{i}}}} \rangle_{\textnormal{\scriptsize{init}}} = 0.02$. The different colours represent the average neutral fraction in the interval we integrate over and the values are represented in the colour bar in the right panel.  Right: As before, but for total hydrogen column density. The dashed black line indicates the column density corresponding to the mean density that we would expect in this velocity interval. }
  \label{front_edge}
\end{figure*}

\begin{figure*}
\includegraphics[trim={1.5cm 12.5cm 1.5cm 9.5cm},clip,width=2.1\columnwidth]{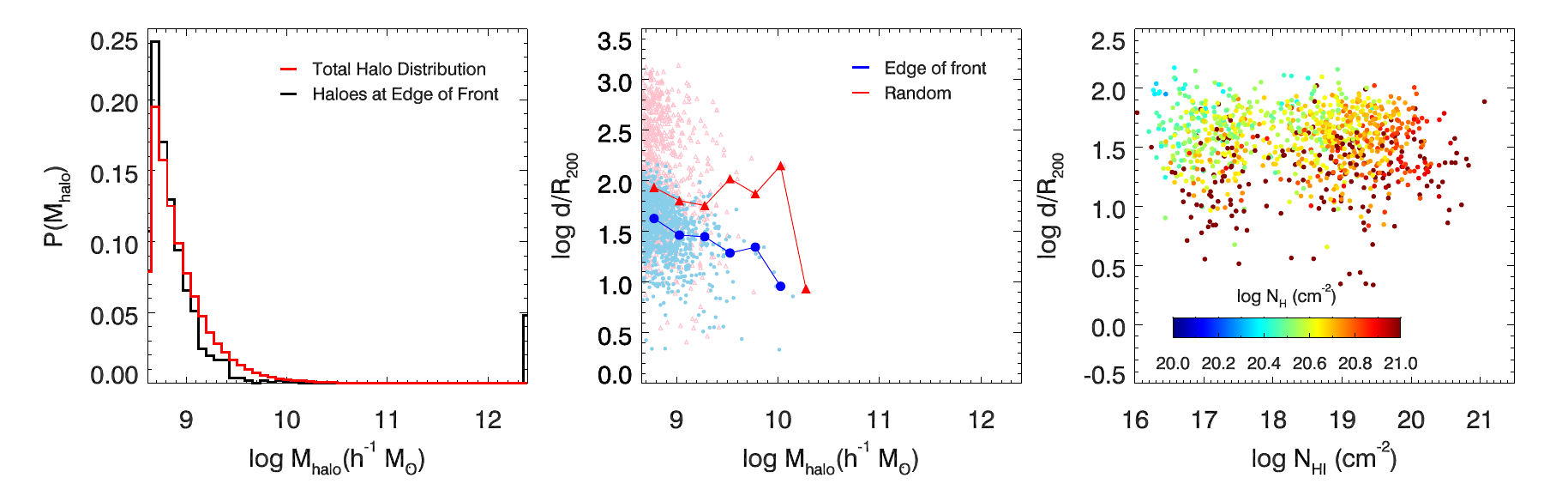}
\caption{Left: A histogram of the masses of the halo closest to the edge of the ionization front  after $t_{\textnormal{\scriptsize{Q}}} = 1$ Myr in our overdense region (black line) and the masses of the dark matter haloes found in our simulation (red line). We find that the lower mass haloes tend to be found nearest to the ionization front. Middle: The distance from the edge of the ionization front to the nearest halo, $d$, normalised by the virial radius, $R_{200}$, against the mass of that halo. The blue line is the median value. The red line is the median value of data measuring the distance to the nearest halo for a random point in our simulation. The contribution from the host halo is not seen because, in that case, $d = 0$. Right: The distance from the edge of the front to the nearest halo $d$, normalised by the virial radius, $R_{200}$, against the \ion{H}{i} column density in a $\pm 500$ km s$^{-1}$ range around the edge of the ionization front after $t_{\textnormal{\scriptsize{Q}}} = 1$ Myr in our overdense region. The points are coloured by the total hydrogen column density.}
  \label{halo_pos}
\end{figure*}

In the top three panels of Figure \ref{fhi_regions}, we show the neutral hydrogen fraction of the gas in the three regions for three of the initial average neutral fractions we consider after $t_{\textnormal{\scriptsize{Q}}} = 1$ Myr at $z=7$. In the cases with  $\langle f_{\textnormal{\scriptsize{\ion{H}{i}}}} \rangle_{\textnormal{\scriptsize{init}}} = 0.002$ and 0.02 (top and second panels), there is a large scatter along different sightlines in the distance the ionization front has travelled, with expansion preferentially into the less dense regions. This is in contrast to the ionization front modelled by \citet{maselli2007} at $z=6.1$, whose simulated \ion{H}{ii} region does not show strong deviations from a spherical ionization front. This is perhaps because our higher-resolution simulations are resolving smaller, dense clumps of neutral gas that produce a shadowing effect, although this effect may become less pronounced if we included the effect of scattered photons from recombinations. In the third panel of Figure \ref{fhi_regions}, we show the case with $\langle f_{\textnormal{\scriptsize{\ion{H}{i}}}} \rangle_{\textnormal{\scriptsize{init}}} = 1.0$. Here we find a more isotropic expansion of the ionization front into the surrounding IGM, with little scatter in the radius of the ionization front and a generally smaller \ion{H}{ii} region. After 5 Myr (bottom panel), although the size of the \ion{H}{ii} region and its scatter have increased, the ionized region is still small compared to the top two panels.

The three columns of Figure \ref{fhi_regions} show our average (left), intermediate (middle) and overdense (right) regions. For each of the initial ionization states, the three regions display a similar maximum distance that the front has travelled and also a similar scatter. This is also shown in the right panel of Figure \ref{haloes}, where we plot the median neutral fraction of all sightlines for our three density fields after the quasar has been on for 1 Myr in the simulation with initial volume-weighted neutral fraction $\langle f_{\textnormal{\scriptsize{\ion{H}{i}}}} \rangle_{\textnormal{\scriptsize{init}}} = 0.02$. Up to a neutral fraction $\log \, f_{\textnormal{\scriptsize{\ion{H}{i}}}} \approx -5$, there is little difference between the three regions but beyond this point, we find that the extent of the \ion{H}{ii} region increases as the density of the environment decreases. After 10 Myr, we find that the ionization front has exited the box and all three regions display similar median neutral fractions.

In the left panel of Figure \ref{front_edge}, we plot the positions and masses of the haloes found in a slice of our simulation the same thickness as the slice over which we project the neutral fraction. Here we use a simulation of our overdense region with initial volume-weighted neutral fraction $\langle f_{\textnormal{\scriptsize{\ion{H}{i}}}} \rangle_{\textnormal{\scriptsize{init}}} = 0.02$ at  $t_{\textnormal{\scriptsize{Q}}} = 1$ Myr, as the ionization front is still contained within our box at this time and for this initial neutral fraction we see a large scatter in the distance the front has travelled. We find that sometimes the progression of the ionization front appears to be impeded by the presence of a halo. This is further investigated in the middle panel of Figure \ref{front_edge}, where we plot the \ion{H}{i} column density found at the edge of the ionization front against the distance that the front has travelled, coloured by the average neutral fraction. We define the edge of the front to be where the smoothed neutral fraction becomes larger than $\log f_{\textnormal{\scriptsize{\ion{H}{i}}}} = -4$. We find that the absorbers slowing the progress of the ionization front tend to be set back from the edge of the front and that we need to integrate over a wide distance of $\pm$500 km s$^{-1}$ (or $\pm$0.6 pMpc, ignoring the contribution of peculiar velocities) around the edge of the ionization front to capture their contribution to the column density. These absorbers have \ion{H}{i} column densities in the range $\log N_{\textnormal{\scriptsize{\ion{H}{i}}}} (\textnormal{cm}^{-2}) = 16-21$. The self-shielding prescription results in a smooth increase of  the neutral fraction of the absorbers. 

We have also looked at the total hydrogen column densities of these absorbers and find that the absorbers with the highest column densities tend to be in the central 2 pMpc of the region, surrounding the host halo. Outside this region, we find hydrogen column densities in the range $\log  N_{\textnormal{\scriptsize{H}}} (\textnormal{cm}^{-2}) = 20.5-21$. At a given distance from the QSO the absorbers with the highest neutral fraction have the highest total column densities, as expected. For comparison, we plot the column density we would expect in our interval at the mean density for $z=7$, which brackets the lower end of the scatter in column density that we measure. 

To investigate  the mass of the dark-matter haloes associated with the edge of the ionization front, we found the halo closest to the point along our sightline where we define the ionization front to end. We find that the closest  dark matter haloes are preferentially  of low-mass ($<10^{9.5} M_{\odot}$, left panel of Figure \ref{halo_pos}). The distribution  thereby roughly tracks the distribution of haloes as a function of mass found in our simulation with a moderate  enhancement at the low-mass end. There are also a significant number of sightlines whose closest halo is the QSO host halo. These correspond to directions in which the ionization front is still trapped within the host halo after $t_{\textnormal{\scriptsize{Q}}} = 1$ Myr.

In the middle panel of Figure \ref{halo_pos}, we examine how the distance from the edge of the front to the closest halo, $d$, changes as a function of halo mass. We normalise this distance by the virial radius $R_{200}$, which we define as the point where the density is equal to 200 times the mean enclosed density. We also choose random points within our simulation and found the closest dark matter halo. In this case, the median distance to the closest halo was slightly larger than this distance from the haloes to the ionization front, but we find a much larger scatter in the measured distances to the nearest halo. There is therefore only a rather loose but still significant association of  the location of the ionization front and dark matter haloes, and this will depend both on where we define the point where the ionization front end and how we choose the halo that is associated with the edge of the ionization front. We have also checked how the \ion{H}{i} and total column density of the absorber scales with the distance to the nearest halo $d$ (right panel of Figure \ref{halo_pos}), but find no strong trend for either quantity.

\subsection{Synthetic Spectra}

\begin{figure*}
\includegraphics[trim={1.5cm 12.5cm 1.5cm 8cm},clip,width=2.1\columnwidth]{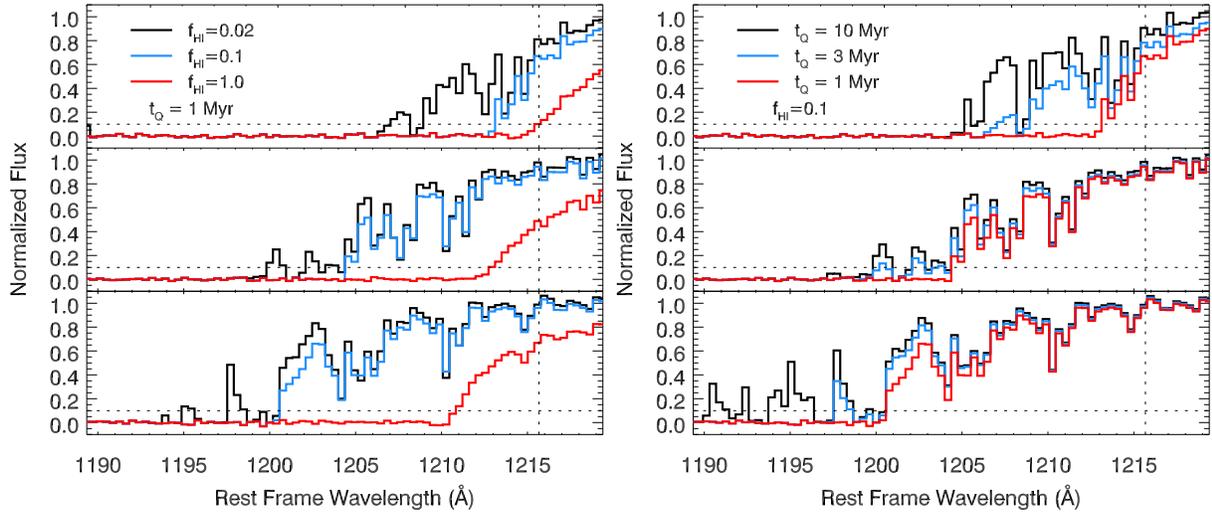}
\caption{Left: Simulated spectra through the overdense region for initial neutral fraction $ \langle f_{\textnormal{\scriptsize{\ion{H}{i}}}} \rangle_{\textnormal{\scriptsize{init}}} = (1.0,0.1,0.02)$ at $t_{\textnormal{\scriptsize{Q}}} = 1$ Myr. Right: Simulated spectra for initial neutral fraction $\langle f_{\textnormal{\scriptsize{\ion{H}{i}}}} \rangle_{\textnormal{\scriptsize{init}}} = 0.1$ at $t_{\textnormal{\scriptsize{Q}}} = (1,3,10)$ Myr. Each row displays a spectrum along a different sightline. The horizontal dotted line marks the point where the transmitted flux is equal to 10 per cent and the vertical dotted line marks the position of rest-frame Lyman-$\alpha$. The spectra have been rebinned onto pixels width 0.4 \AA \, in the rest frame.}
  \label{spectra}
\end{figure*}

We used our simulation to produce mock spectra, from which we measured the QSO near-zone sizes  and compared to the observed near-zones. The mock  \lya\ spectra were constructed with rest-frame wavelength $\lambda_{\alpha} = 1215.67$ \AA, damping constant $\Lambda_{\alpha} = 6.265 \times 10^8 \, \textnormal{s}^{-1}$ and oscillator strength $f_{\alpha} = 0.4164$. The peculiar velocity and temperature of the gas were taken into account. To make our spectra more realistic, we added noise and convolved them with a Gaussian instrument profile with a FWHM of 240 km s$^{-1}$.

Figure \ref{spectra} shows examples of spectra constructed along three lines of sight in our overdense region at $z=7$, varying both the initial average volume-weighted neutral fraction and quasar lifetime. We find that, as expected from Figure \ref{fhi_regions}, the Ly$\alpha$ transmission at fixed lifetime extends for longer distances in a more ionized IGM. The spectra constructed from simulations with  $\langle f_{\textnormal{\scriptsize{\ion{H}{i}}}} \rangle_{\textnormal{\scriptsize{init}}} = 0.02$ and 0.1 show broadly similar transmission inside the \ion{H}{ii} region, while the spectra from the $\langle f_{\textnormal{\scriptsize{\ion{H}{i}}}} \rangle_{\textnormal{\scriptsize{init}}} = 1.0$ display strong damping wings. When the initial neutral fraction is kept constant at $\langle f_{\textnormal{\scriptsize{\ion{H}{i}}}} \rangle_{\textnormal{\scriptsize{init}}} = 0.1$ and the QSO lifetime varied (right panel of Figure \ref{spectra}), we see that along some sightlines (middle and bottom panels) the front initially grows  rapidly but  the growth saturates already after 1 Myr. When there is a dense absorber along the sightline (top panel), saturation of the radius of the front is not yet reached  even after 10 Myr. In this case, the absorber has a total hydrogen column density $\log N_{\textnormal{\scriptsize{H}}} (\textnormal{cm}^{-2}) = 21.5$ (integrating over the same velocity interval as in Section \ref{sec:fronts}). The density begins to rise rapidly after a distance 3.4 pMpc from the source (which corresponds to the edge of the ionization front) before peaking at 3.7 pMpc.

We determined  the extent of the near-zone along each line of sight, which, following \citet{fan2006}, we defined as the distance from the source to the point where the transmitted flux drops below 10 per cent for the first time after smoothing our spectra with a box car window of width 20 \AA\ in the observed frame. This smoothing reduces the effect of noise and intervening absorption lines.

\subsection{Near-Zones vs. Proximity Zones: Analytic Models}

It is important to note that the size of the \ion{H}{ii} region is not necessarily the size of the near-zone that will be observed. \citet{bolton2007a} summarise the analytical solution for the radius of an ionization front moving into the IGM ($R_{\textnormal{\scriptsize{ion}}}$). For an source emitting $\dot{N}$ ionizing photons per unit time, the rate of expansion of the \ion{H}{ii} region is given by
\begin{equation}
\frac{\textnormal{d}R_{\textnormal{\scriptsize{ion}}}}{\textnormal{d}t} = \frac{\dot{N} - \frac{4}{3} \pi R_{\textnormal{\scriptsize{ion}}} \alpha_{\textnormal{\scriptsize{\ion{H}{ii}}}} n_{\textnormal{\scriptsize{H}}}^2}{4 \pi R_{\textnormal{\scriptsize{ion}}}^2 f_{\textnormal{\scriptsize{\ion{H}{i}}}} n_{\textnormal{\scriptsize{H}}}},
\end{equation}
where $f_{\textnormal{\scriptsize{\ion{H}{i}}}}$ is the neutral hydrogen fraction, $n_{\textnormal{\scriptsize{H}}}$ is the total hydrogen number density and $\alpha_{\textnormal{\scriptsize{\ion{H}{ii}}}}$ is the recombination coefficient for ionized hydrogen. Note that we do not use the case A recombination rate as in \cite{bolton2007a}, but the case B recombination rate as we are using the ``on-the-spot'' approximation for recombination. This difference is discussed further in Section \ref{env}. Solving for $R_{\textnormal{\scriptsize{ion}}}$, and assuming that the lifetime of the source $t_{\textnormal{\scriptsize{Q}}}$ is much smaller than the timescale for recombination $t_{\textnormal{\scriptsize{rec}}} = (n_{\textnormal{\scriptsize{H}}}\,\alpha_{\textnormal{\scriptsize{\ion{H}{ii}}}})^{-1}$, leads to the solution
\begin{equation}
\begin{split}
R_{\textnormal{\scriptsize{ion}}} &=\frac{1.48}{(\Delta f_{\textnormal{\scriptsize{\ion{H}{i}}}})^{1/3}}\bigg(\frac{\dot{N}}{1.3 \times 10^{57} \, \textnormal{s}^{-1}}\bigg)^{1/3}\bigg(\frac{t_{\textnormal{\scriptsize{Q}}}}{10^6 \, \textnormal{yr}}\bigg)^{1/3} \\
&\times\bigg(\frac{1+z}{8}\bigg)^{-1} \, \textnormal{pMpc}.
\end{split}
\end{equation}

Due to residual neutral hydrogen inside the \ion{H}{ii} region, the absorption can become saturated closer to the source. By choosing the $f_{\textnormal{\scriptsize{\ion{H}{i}}}}$ that corresponds to the limiting optical depth and assuming that the hydrogen gas beyond the ionization front is in ionization equilibrium, the largest observable near-zone size, $R_{\alpha}^{\textnormal{\scriptsize{max}}}$, can be defined by
\begin{equation}
\begin{split}
R_{\alpha}^{\textnormal{\scriptsize{max}}} & \simeq \frac{1.86}{\Delta_{\textnormal{\scriptsize{lim}}}}\bigg(\frac{\dot{N}}{1.3 \times 10^{57} \, \textnormal{s}^{-1}}\bigg)^{1/2}\bigg(\frac{T}{2 \times 10^4 \textnormal{K}}\bigg)^{0.35} \\
& \times
\bigg(\frac{\tau_{\textnormal{\scriptsize{lim}}}}{2.3}\bigg)^{1/2}\bigg(\frac{\alpha^{-1}[\alpha+3]}{3}\bigg)^{-1/2}\bigg(\frac{1+z}{8}\bigg)^{-9/4} \, \textnormal{pMpc},
\end{split}
\label{rmax}
\end{equation}
where $\Delta_{\textnormal{\scriptsize{lim}}}$ is the overdensity corresponding to the limiting neutral fraction of hydrogen, $T$ is the temperature, $\tau_{\textnormal{\scriptsize{lim}}}$ is the optical depth detection limit and $\alpha$ is the spectral index of the QSO. Note that this scales as $\dot{N}^{1/2}$ rather than $\dot{N}^{1/3}$ as before and that there is no dependence on $f_{\textnormal{\scriptsize{\ion{H}{i}}}}$ or $t_{\textnormal{\scriptsize{Q}}}$. We confirmed this scaling using our radiative transfer simulations for luminosities varying over a range of 1 dex below our fiducial luminosity.

\subsection{The Sizes of Simulated Near-Zones}
\label{env}

\begin{figure*}
\includegraphics[trim={1.5cm 12.5cm 1.5cm 5cm},clip,width=2.1\columnwidth]{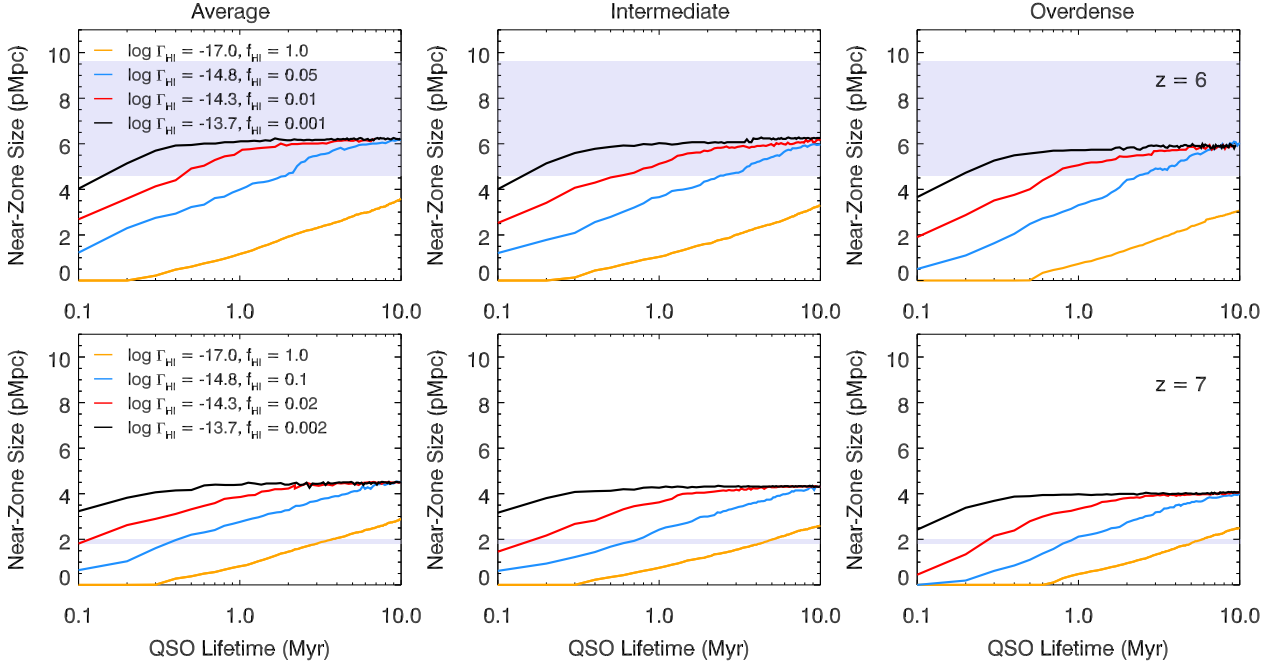}
\caption{Evolution of QSO near-zone size with time for the three different density fields. Left: average, middle: intermediate, right: overdense. Top: $z=6$, bottom: $z=7$. In each panel we plot a range of different initial neutral fractions. The solid line in each case is the median near-zone size we measure for the system. The shaded region in the top panels shows observed near-zone sizes, rescaled to match the luminosity of our source, in the range $5.9<z<6.1$ \citep{carilli2010} and in the bottom panels it shows the estimate for the near-zone size of the $z=7.085$ QSO ULAS J1120+0641.}
 \label{tq_all}
\end{figure*}

\begin{figure*}
\includegraphics[trim={1.5cm 12.5cm 1.5cm 9.5cm},clip,width=2.1\columnwidth]{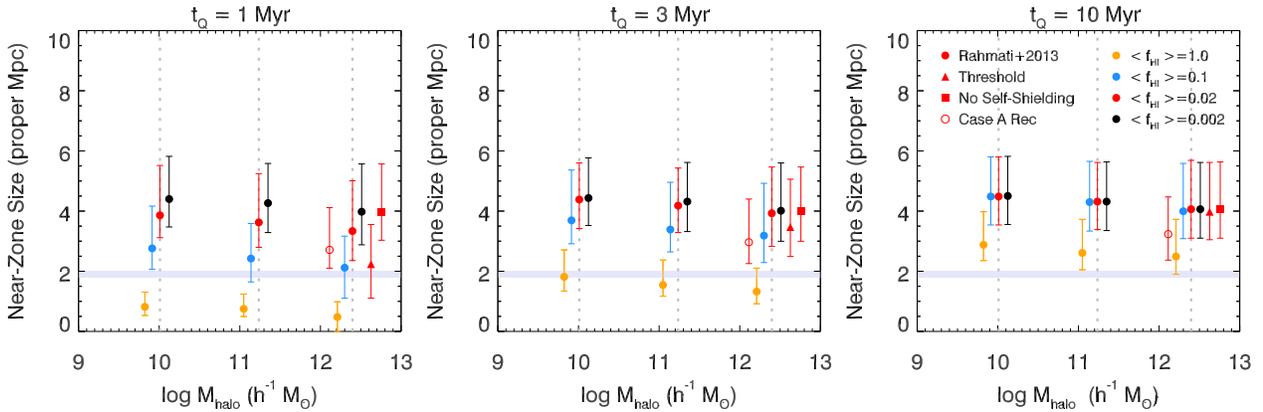}
\caption{Median near-zone size as a function of mass of the host halo at $z=7$. The shift in the points around the masses of the three haloes is for display purposes only. The dashed grey vertical lines mark the mass of the host halo. The error bars represent the  15$^{\textrm{th}}$/85$^{\textrm{th}}$ percentiles in the near-zone sizes. No strong trend of near-zone size with the mass of the host halo is visible. The shaded region shows the estimate for the near-zone size of the $z=7.085$ QSO ULAS J1120+0641.}
  \label{mass}
\end{figure*}

We investigate the dependence of the near-zone size on the mass of the host halo by following the evolution of the median near-zone size with time in our three different environments at $z=6$ and 7 (Figure \ref{tq_all}). Note that from this point on, due to the computational costs running the radiative transfer simulations to 10 Myr, we only perform the radiative transfer on one octant of our simulation box. We  find that once the ionization front has moved beyond the point where the clustering around host halo boosts the density of the IGM, there is only a very weak trend of decreasing near-zone size with increasing  mass of the host halo. The other point of interest is the manner in which the near-zone sizes grow. There is a period of initial rapid growth, followed by a levelling-off in which there is little evolution in the median near-zone size of the system. The period of rapid growth is the time for which $R_{\textnormal{\scriptsize{ion}}} < R_{\alpha}^{\textnormal{\scriptsize{max}}}$. After this point, we find that the median near-zone size and scatter in the near-zone sizes are independent of the initial neutral fraction. Due to the size of our simulation, at $z=6$ approximately 10 per cent of our sightlines have near-zones that extend beyond the size of our box. In this case, we take the size of the box to be a lower limit to the near-zone size. In Appendix \ref{sec:feedback}, we investigate the effect of including thermal feedback (from AGN and supernovae) in our hydrodynamical simulations and find that although this does not change our results significantly for $t_{\textnormal{\scriptsize{Q}}} > 1$ Myr, we do find larger near-zone sizes for QSO lifetimes shorter than this due to a bubble of hot gas centred on the AGN. We find also that there is a natural decrease in the distance at which the near-zone sizes become saturated with increasing redshift. As shown by equation \ref{rmax}, such a decrease is expected due to the increase of the mean baryonic density with redshift,  as $\Delta_{\rm lim}$ and the IGM  temperature are probably only varying slowly. We see also a small decrease in the scatter with increasing redshift.

In Figure \ref{mass}, we plot the near-zone sizes we measure after $t_{\textnormal{\scriptsize{Q}}} = 1,3$ and 10 Myr as a function of the mass of the host halo at $z=7$. Results are shown for a range of initial average neutral fractions. Again, we find only a weak correlation between the near-zone size and the density of the three regions. This can be easily understood  by looking at the left panel of Figure \ref{haloes}, which shows the mean overdensity contained in a sphere of a given radius centred around the host halo. When the size of the observed near-zones are small enough, the matter overdensity around the host halo has not yet returned to the cosmic mean and the effect of the immediate environment  of the host  halo is imprinted  in the distribution of near-zone sizes. However, due to the large scatter in near-zone sizes found in all three cases,  it will  be difficult to use near-zone sizes to constrain the mass of the host halo of the QSO.

For our overdense halo, we also investigate how our choice of self-shielding model effects the sizes of the near-zones we measure. For the same photoionization rate $\log \Gamma_{\textnormal{\scriptsize{\ion{H}{i}}}}$ (s$^{-1}$) = -14.3, we compare our fiducial model, taken from \citet{rahmati2013}, with a simple threshold self-shielding model where all gas with overdensity above some threshold $\Delta_{\textnormal{\scriptsize{ss}}}$ \citep{schaye2001} is assumed to be neutral \citep[see also][]{mhr2000,furlanetto2005,bolton2011}.   We find that in the simple threshold model, the Lyman limit systems become more effective at slowing the progression of the ionization front. This results in a smaller \ion{H}{ii} region and near-zone sizes. However, the distance at which the near-zone sizes saturate is similar for both self-shielding models and for the case of no self-shielding at all. Note also that the largest effect of changing the self-shielding model is the amount of neutral gas available for the same photoionization rate. The difference between the threshold self-shielding model and the prescription of  \citet{rahmati2013} increases the average volume weighted neutral hydrogen fraction by an order of magnitude and neglecting self-shielding entirely decreases the average volume weighted neutral hydrogen fraction by an order of magnitude.

We also check the effect of switching from case B to case A recombination rates.  We find that the near-zone sizes saturate at a distance of 3 pMpc from the host halo, closer to the value obtained in \citet{bolton2011} than what we find in our models using case B recombination. However, we still do not include the diffuse radiation that would result from a full treatment of recombinations. The change in recombination rate does not make a large difference for the smaller near-zones we measure and only becomes important along sightlines where the ionization front has travelled a large distance when the number of recombinations along the sightline becomes significant. This therefore also has the result of reducing the scatter slightly. 

\subsection{Red Damping Wings in the Simulated Spectra}

\begin{figure*}
\includegraphics[trim={1.5cm 12.5cm 1.5cm 5cm},clip,width=2.1\columnwidth]{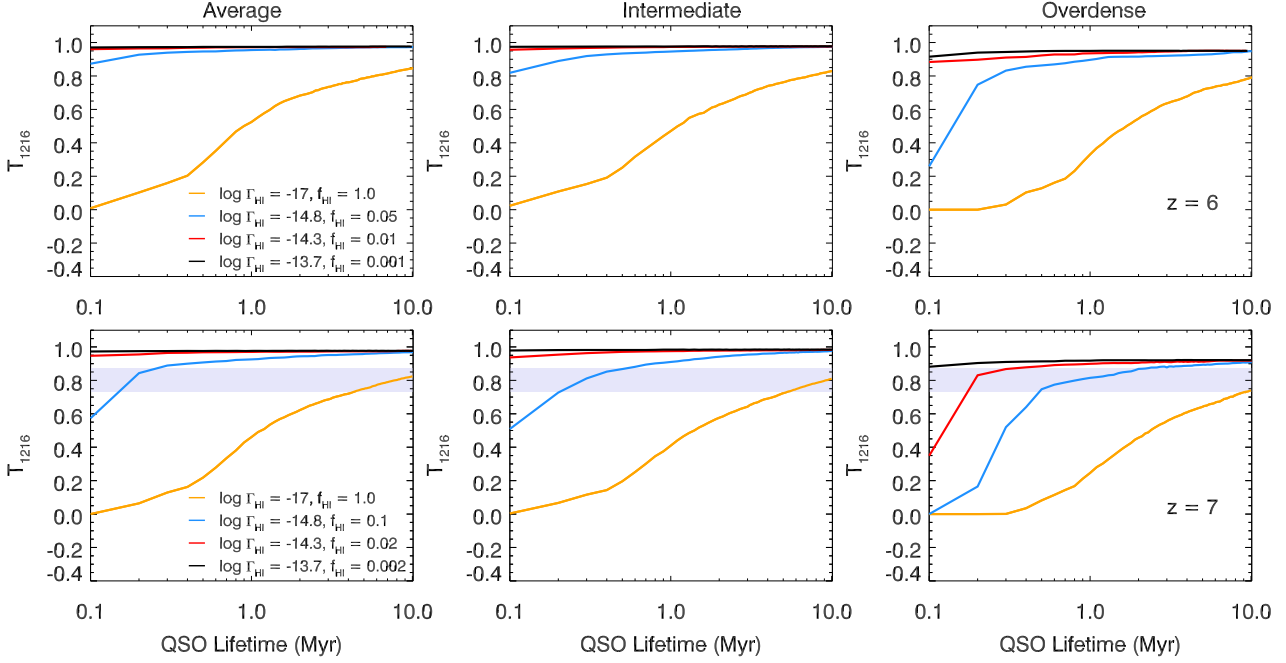}
\caption{Evolution of transmission at Ly$\alpha$ with time for the three different density fields. Left: average, middle: intermediate, right: overdense. Top: $z=6$, bottom: $z=7$. In each panel we plot a range of different initial neutral fractions. The solid line in each case is the median transmission at 1216 \AA\ we measure. More sightlines display a red damping wing in the overdense region than in the less dense regions. The shaded region shows the estimate for the transmission at Ly$\alpha$ of the $z=7.085$ QSO ULAS J1120+0641.}
 \label{t1216_all}
\end{figure*}

\begin{figure*}
\includegraphics[trim={1.5cm 12.5cm 1.5cm 9.5cm},clip,width=2.1\columnwidth]{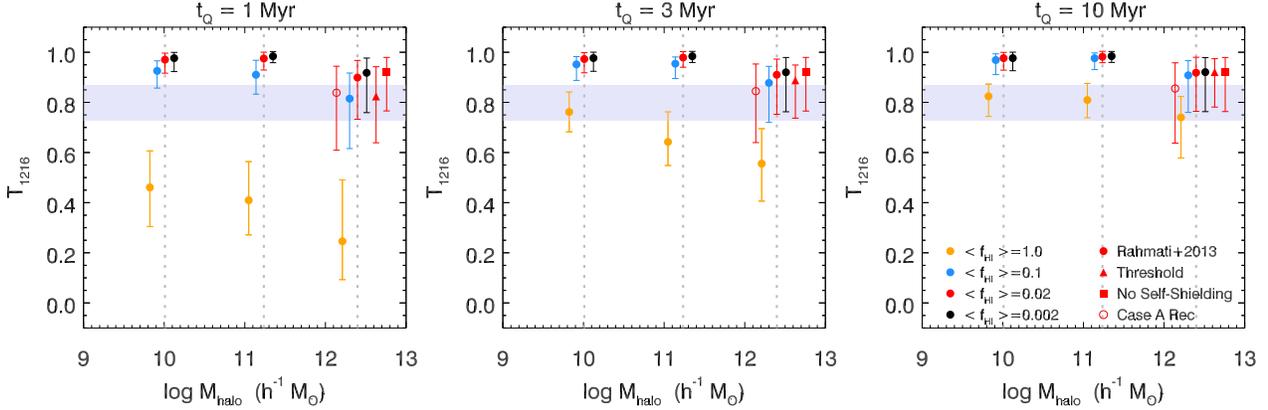}
\caption{Median transmission at Ly$\alpha$ as a function of mass of the host halo at $z=7$. The shift in the points around the masses of the three haloes is for display purposes only. The dashed grey vertical lines mark the mass of the host halo. The error bars represent the 15$^{\textrm{th}}$/85$^{\textrm{th}}$ percentile scatter in the near-zone sizes. The shaded region shows the estimate for the transmission at Ly$\alpha$ of the $z=7.085$ QSO ULAS J1120+0641.}
  \label{mass_tr}
\end{figure*}

The second observable that has been used to quantify the impact of a still largely neutral IGM on the spectra of QSOs is the transmission at the rest-frame wavelength of Ly$\alpha$. In the presence of high densities of neutral hydrogen, even flux redward of Ly$\alpha$ can be attenuated producing a red damping wing. The strength of this damping wing can then be used to place constraints on the neutral fraction of the IGM \citep{miraldaescude1998,mesinger2004,wyithe2004,mesinger2008,bolton2011,mortlock2011,schroeder2013}. 

In Figure \ref{t1216_all}, we plot the transmission at the Ly$\alpha$ rest-frame wavelength as a function of time for our three regions. We measure this transmission after smoothing our spectra to the same resolution at which we measure the near-zone sizes. We find that, as expected, the regions with higher average initial neutral fractions display stronger red damping wings. These damping wings are stronger for shorter QSO lifetimes, when the ionization front is still in the higher-density region surrounding the host halo.  

Perhaps more interestingly, we also find that the transmission at Ly$\alpha$ is consistently lower in the overdense regions, for all the initial average neutral fractions that we consider due to the increased number of dense, self-shielded systems. This is displayed more clearly in Figure \ref{mass_tr}. It is perhaps not surprising that we find that the mass of the host halo plays an important role in the strength of the red damping wing and not for the size of the near-zones. In the case of the near-zones, we are probing parts of the IGM with average density close to the mean. The strength of the damping wing is however more sensitive to the environment around the source. This clustering of dense absorbers is largest in the most massive halo we look at (left panel of Figure \ref{haloes}), resulting in a large scatter in the transmission at at Ly$\alpha$ not found in the lower mass haloes we simulate.

In this case, the choice of self-shielding model does not seem to make a large difference. For short QSO lifetimes, we find that the threshold self-shielding model predicts a stronger damping wing. For longer, more realistic lifetimes ($\sim 10$ Myr), the results from the threshold self-shielding model are similar to those predicted by the \citet{rahmati2013} model. We also do not find a large change in the transmission when we use the case A recombination rate.

\section{Comparison with Observations}

\subsection{$z \sim 6$ Near-Zones}

\begin{figure*}
\includegraphics[trim={1.5cm 12.5cm 1.5cm 9.5cm},clip,width=2.1\columnwidth]{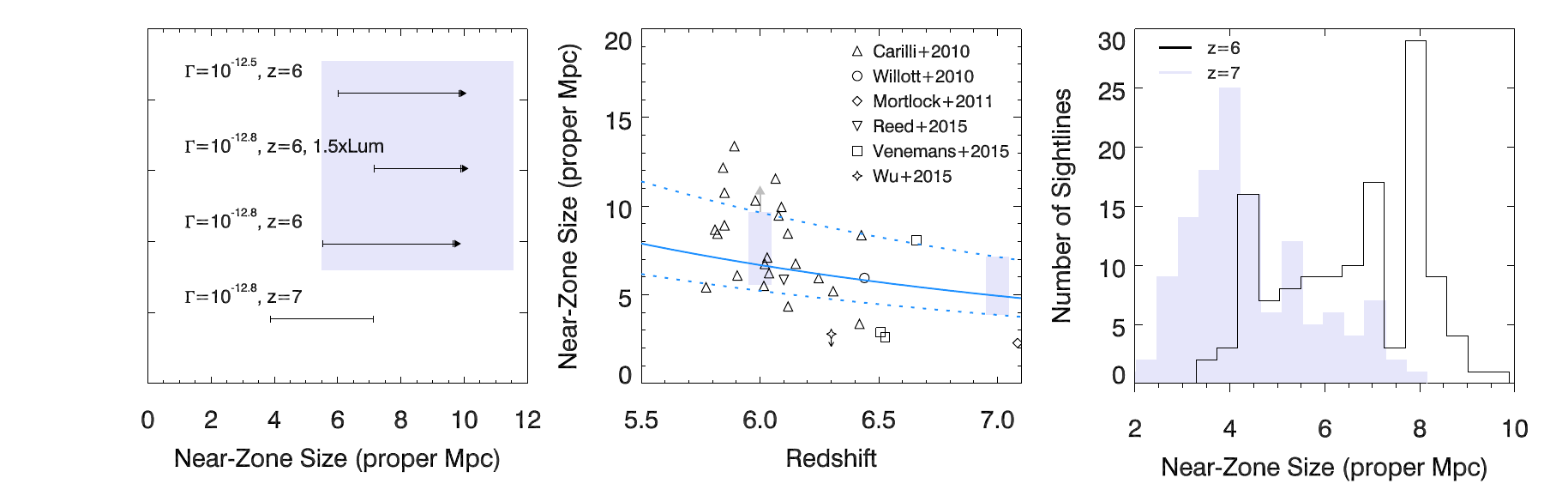}
\caption{Left: QSO near-zone sizes (corrected for differences in luminosity) at $z = 6$ and $z=7$ for background photoionization rates ($\log \Gamma_{\textnormal{\scriptsize{\ion{H}{i}}}}$ (s$^{-1}$) = -12.8, -12.5) and two different QSO luminosities ($\log \dot{N}_{\gamma}  (\textnormal{s}^{-1})$  = 57.1, 57.3) in our overdense region after 10 Myr. The shaded region represents the observed near-zone sizes at $z=6$ taken from \citet{carilli2010}. Middle: Observed near-zone sizes as a function of redshift, scaled to the same luminosity. Plotted in blue is the analytic evolution of near-zone sizes with redshift given by Equation (\ref{rmax}) with $\Delta_{\textnormal{\scriptsize{lim}}}$ at the median and the measured 15$^{\textrm{th}}$/ 85$^{\textrm{th}}$ percentiles of the saturated near-zone sizes at $z=7$. The shaded regions are the 15$^{\textrm{th}}$/85$^{\textrm{th}}$ percentiles of the simulated near-zone sizes at $z=6,7$. Right: Distribution of the near-zone sizes at $z=6$ and 7 in the simulations with $\log \Gamma_{\textnormal{\scriptsize{\ion{H}{i}}}}$ (s$^{-1}$) = -12.8 after 10 Myr for our default luminosity.}
  \label{nz_redshift}
\end{figure*}

In the left panel of Figure \ref{nz_redshift}, we compare the measured near-zone sizes from simulation of our overdense region at $z=6$ to observed near-zone sizes in the range $5.9<z<6.1$ taken from \citet{carilli2010}. Here, we used a background photoionization rate $\log \Gamma_{\textnormal{\scriptsize{\ion{H}{i}}}} (\textnormal{s}^{-1}) = -12.8$, based on measurements of the UV background at $z \sim 6$ by \citet{calverley2011} and \citet{bolton2007a} and $\log \Gamma_{\textnormal{\scriptsize{\ion{H}{i}}}} (\textnormal{s}^{-1}) = -12.5$ to mimic a late reionization scenario \citep[as in ][]{choudhury2014}. We consider two different QSO luminosities, $\log \dot{N}_{\gamma} (\textnormal{s}^{-1}) = 57.1$ and 57.3, which fall in the range of values estimated by \citet{fan2006}. 

For the same quasar luminosity and background photoionization rate, there is an increase in the near-zone size with decreasing redshift. With our fiducial quasar luminosity and a UV background $\log \Gamma_{\textnormal{\scriptsize{\ion{H}{i}}}} = -12.8$, we match the lower end of the observed $z=6$ near-zone sizes remarkably well. This is somewhat in contrast with other work \citep{bolton2007a,wyithe2008} and the different result is most likely due to the difference in the recombination rates used here (case B {\it vs.} case A). We have also run  a simulation where we increased the luminosity of our ionizing source by a factor of 1.5 and find that this increases the near-zone sizes we find with many now larger than the near-zone sizes our box can contain.  In the cases where the near-zone size occurred at a distance larger than the radius of our box, we took the radius of the box as a lower limit.

\subsection{The $z=7.085$ QSO ULAS J1120+0641}

We now want to use our  simulations to revisit the constraints on the ionization state of the IGM around the $z=7.085$ QSO ULAS J1120+0641. Modelling of this kind has previously been undertaken in \citet{bolton2011}. However, the mass of the host halo in that work was a factor of 100 times less massive than the one considered here (although based on section \ref{env} we  expect this to make remarkably little  difference).

Our analytical estimate  described by equation (\ref{rmax}) predicts that $R_{\alpha}^{\textnormal{\scriptsize{max}}} = 1.86/\Delta_{\textnormal{\scriptsize{lim}}}$ proper Mpc (assuming our fiducial values for $\alpha$ and $\dot{N}$).  Assuming that the absorption at $z=7.085$ is due to  overdensities of $\Delta \sim 0.5$ similar to that required to explain the observed near-zone sizes at $z=6$, then we would expect $R_{\alpha}^{\textnormal{\scriptsize{max}}} \sim 3.72$ (albeit  with a large scatter).  \citet{mortlock2011} measure the near-zone of  ULAS J1120+0641 to be 1.9 $\pm$ 0.1 proper Mpc.  This suggests  that the near-zone of ULAS J1120+0641 is unlikely to be due to the  proximity effect and may therefore  be  a useful probe  of a substantially neutral  IGM at $z=7.085$.

In the right panel of Figure \ref{tq_all}, we compare the measured near-zone size of ULAS J1120+0641 (shaded purple region) to the simulations of the overdense region containing a $2.5\times 10^{12} \, M_{\odot}$ halo. We find that  only for simulations with a large  average initial neutral fraction somewhere between $\langle f_{\textnormal{\scriptsize{\ion{H}{i}}}} \rangle_{\textnormal{\scriptsize{init}}} = 0.1-1.0$  can we reproduce the small observed near-zone size.

For  QSO lifetimes long enough to allow the central massive black hole to grow while the QSO is optically bright, we also struggle to match the strength of the damping wing observed in ULAS J1120+0641 except in the case of our most massive host halo or for an initially neutral IGM.  For the simulations centred on our most massive halo, damping wings almost as strong as that of  ULAS J1120+0641 are seen in the 15$^{\textrm{th}}$/ 85$^{\textrm{th}}$ percentiles of all of our models, even the ones with low volume weighted initial neutral hydrogen fraction (e.g., $\langle f_{\textnormal{\scriptsize{\ion{H}{i}}}} \rangle_{\textnormal{\scriptsize{init}}} = 0.02$). The observed red damping wing therefore provides little constraint  on the neutral fraction of the surrounding IGM if ULAS J1120+0641 is indeed hosted by a massive halo.  
 However, note here that we have neglected  the ionizing radiation from Lyman-$\alpha$ emitters that may reside in these systems and weaken the Ly$\alpha$ damping wing \citep{bolton2013}.

Note also that modelling of this kind is further complicated by the fact that, towards the end of reionization, there are expected to be large variances in the neutral fraction along different sightlines \citep[e.g.,][]{becker2015} and that the QSOs  may have been bright  at earlier redshifts and may have already ionized its surroundings \citep{feng2013}.

\subsection{Evolution with Redshift}

In the middle panel of Figure \ref{nz_redshift}, we plot the published near-zone sizes of a selection of quasars with $z > 5.8$ \citep{carilli2010,willott2010a,mortlock2011,reed2015,venemans2015,wu2015}. Near-zone sizes have been corrected for luminosity as $R_{\textnormal{\scriptsize{NZ,corr}}}=10^{0.4(27+M_{\textnormal{\scriptsize{1450,AB}}})/2}R_{\textnormal{\scriptsize{NZ}}}$. Note that near-zone sizes in the literature have been often corrected assuming  that the near-zone size scales with the ionizing luminosity  as $\dot{N}^{1/3}$, as expected when  the near-zone size traces the edge of the \ion{H}{ii} region, not $\dot{N}^{1/2}$ which is the (more likely correct) scaling expected for  saturated near-zone sizes. The recently reported rather large near-zone size measured by \citet{wu2015} decreases substantially when rescaled for the luminosity of this very bright QSO. We plot this particular near-zone size as an upper-limit, as  \citet{wu2015} have used  a  definition of  near-zone size  for their measurement that is different from  the normally used definition from   \citet{fan2006}.

For comparison, we also plot the values at the 15$^{\textrm{th}}$/85$^{\textrm{th}}$ percentiles of near-zone sizes we measure in our simulations at $z=6$ and 7, also rescaled for luminosity (with an assumed $M_{\textnormal{\scriptsize{1450,AB}}}$ to match ULAS J1120+0641). Here we use a highly ionized IGM (with background photo-ionization rate $\log \Gamma_{\textnormal{\scriptsize{\ion{H}{i}}}}$ (s$^{-1}$) = -12.8, to try and make a fair comparison to the observed $z=6$ near-zones). For this $\Gamma_{\textnormal{\scriptsize{\ion{H}{i}}}}$, we find that the near-zone sizes saturate quickly. Here we plot the scatter of a simulation  at $t_{\textnormal{\scriptsize{Q}}} = 10$ Myr, but in this regime we no longer expect any dependence on  $t_{\textnormal{\scriptsize{Q}}}$ (Figure \ref{tq_all}).  We also plot the analytic solution for the evolution of saturated near-zone sizes with redshift (Equation \ref{rmax}), where we have selected $\Delta_{\textnormal{\scriptsize{lim}}}$ to fit the median near-zone size and the scatter we measure at $z=7$ [$\Delta_{\textnormal{\scriptsize{lim}}}$ = (0.26, 0.38, 0.48) for the 15$^{\textrm{th}}$ percentile, median and the 85$^{\textrm{th}}$ percentile].

The analytic solution tuned to fit our results at $z=7$ also agrees  well with our results at $z=6$. Note that the evolution we predict is shallower than the linear fits to the near-zone sizes from \citet{carilli2010} and \citet{venemans2015} perhaps  suggesting that the IGM indeed becomes substantially more neutral at $z>6.3$ than predicted by the evolution of the mean density alone. We also note that the scatter at $z=6$ will in reality be somewhat larger (indicated by the grey arrow in the plot), as we have used the size of our box as a lower limit in cases where we do not find a near-zone size along a sightline. At $z > 6.5$ there are several near-zones that fall outside of our predicted scatter. These near-zones may represent unsaturated cases, where the size is still following the radius of the \ion{H}{ii} region. For realistic QSO lifetimes $t_{\textnormal{\scriptsize{Q}}}  \gtrsim 1$ Myr, Figure \ref{tq_all} suggests that these QSO  may reside in regions of the IGM with $\langle f_{\textnormal{\scriptsize{\ion{H}{i}}}} \rangle_{\textnormal{\scriptsize{init}}}  \gtrsim 0.1$ and lower photo-ionization rate than we assumed in Figure \ref{tq_all}. 

The observations at $z \sim 6$ contain near-zones with sizes $\ga 10$ pMpc which our simulations cannot reproduce due to the  limited box size of our simulation.  The distribution of near-zone sizes at $z=6$ shown in the right panel of Figure \ref{nz_redshift} has a large peak at 8 pMpc. This is the result of the lower limits we are assuming for sightlines where no near-zone is found. For a larger simulated region some of our sightlines should  produce near-zones that would match the observations. However, these would be the outliers in our distribution and it is also possible that defining the point where the near-zone ends at $z \sim 6$ becomes ambiguous due to patches of transmitted flux in the Ly$\alpha$ forest \citep{bolton2007a}. 

\subsection{Caveats and Limitations}

There are obviously still a significant number of uncertainties and limitations in our modelling that should be taken into account when considering  our results. First, the resolution of our hydrodynamical simulations is still somewhat below what would ideally  required to resolve the Ly$\alpha$ forest \citep{bolton2009} at $z > 5$. Our maximum resolution  was dictated by the large simulation region necessary  to follow the evolution of the ionization front over a realistic quasar lifetime. For the simulation box to still contain observed near-zone sizes at $z>6.5$ , the chosen resolution was the best that was still computationally feasible to run. As we have discussed, the largest observed near zone sizes at $z\sim 6$ still exceed the size of the region we have simulated here by up to a factor 1.5. We have investigated the effects of changing the resolution in Appendix \ref{sec:restest}.
  
As stated in Section \ref{sec:model_qso}, our radiative transfer post-processing only evolves the temperature and ionization state of the gas, but neglects the evolution of the density field as well as the coupling of the radiation to the density field. However, as discussed earlier the timescales that we are considering here are short compared to the timescales over which this matters. The speed of the ionization front in our case is close to the speed of light, and therefore many orders of magnitude larger than the sound speed in the IGM. Another possible effect we neglected is the thermal pressure due to the heating provided by the passage of the ionization front. But, once again, adiabatic expansion of the gas is negligible over the timescales that we are considering. Also, as we are using the ``on-the-spot'' treatment of recombination radiation, we assume that ionizing photons produced by hydrogen and helium recombinations are re-absorbed instantaneously in the same computational cell. The inclusion of the diffuse radiation from recombinations would likely reduce the ability of dense absorbers to effectively shadow the ionization front (as seen in Figure \ref{fhi_regions}) and somewhat reduce the scatter in the near-zone sizes we model [see \citet{cantalupo2011} for a detailed discussion of recombination radiation effects on quasar \ion{H}{II} regions].

We furthermore only model the initial ionization state of the IGM using a homogeneous UV background. This certainly oversimplifies the previous evolution of the IGM before the QSO turns on for a single episode of constant luminosity as reionization is a highly inhomogeneous process, with individual galactic sources pre-ionizing inhomogenously the surrounding IGM to different levels and the QSO likely to have a complex light curve. An enhanced contribution to the photoionization rate  from galaxies close to the QSO has been previously suggested to increase  near-zone sizes at $z=6$ \citep{bolton2007a,wyithe2008,maselli2009}. However, this simple model for the UV background should not change our predictions for the evolution of saturated near-zone sizes with redshift (middle panel of Figure \ref{nz_redshift}), as the distance at which the near-zone sizes become saturated at is independent of the ionization state of the gas. It will also not change our results at $z=7$, as we find it difficult to match the observed near-zone of  ULAS J1120+0641 even for an IGM that is initially completely neutral.

\section{Summary and Conclusions}

We have presented here results from our detailed modelling of quasar near-zones  at $z=6$ and 7. We have performed  zoom simulations of three dark matter haloes from the Millennium simulation, which differ in mass by more than a  factor of 100, and post-processed these with the 3D radiative transfer code \textsc{radamesh}, exploring a range of initial neutral fractions representing different reionization histories.  We placed a source of the same luminosity as expected for the $z=7.085$ QSO ULAS J1120+0641 in each halo.

The spatial extent of the ionization fronts in our simulations  shows a large scatter with direction for fixed QSO lifetime. The location of the edge of ionization front appears to have a loose association with  low-mass dark matter haloes ($>10^{9} M_{\odot}$) hosting self-shielded absorbers. We further find that the size  of the near-zones of bright high-redshift QSOs  depends only very weakly on the  mass of the QSO host halo. For  realistic QSO lifetimes ($> 1$ Myr), the ionization front has moved out into the general IGM where the average densities have returned to close to the mean. 
 
Our simulations reproduce the observed range of $z\sim6 -7$ near-zone sizes very well once we take into account  the limitations due the size of our  simulated region. Even when the extent of the near-zone sizes in our simulation has become saturated due to residual neutral hydrogen in the \ion{H}{ii} region, we still find a large scatter in the near-zone sizes along different sightlines around the host halo. The scatter of the observed flux due to variations in the density field can explain a lot of  the scatter in the observed near-zone sizes without the need  to invoke different volume averaged neutral hydrogen fractions along different lines of sight or anisotropic emission. 

We have  presented here the first simulations of QSO near-zones around a host halo with a mass as high as expected for  ULAS J1120+0641. We find that, for a QSO lifetime of 10 Myr, the observed near-zone size of ULAS J1120+0641 falls just within the lower end of our 15$^{\textrm{th}}$/85$^{\textrm{th}}$ percentile scatter for a IGM that is initially completely neutral at $z=7$. This may be indicative of the presence of a high \ion{H}{i} column density absorber lying close to the quasar or  may suggest that the duration of the optically bright phase in ULAS J1120+0641 is $<10$ Myr and thus too short for its supermassive black hole to grow significantly. Alternatively, this  may   cast doubt on the robustness of the measured size of this particular near-zone.  We further  caution that for  massive QSO host haloes, the transmission at the Ly$\alpha$ rest-frame wavelength can be low even if the IGM is generally highly ionized.

Current and upcoming surveys will provide new observations of QSOs and their near-zone sizes which, in conjunction with alternative observations, should allow us to further improve constraints on the neutral fraction of the  IGM at the end of reionization, as well as on the relative contribution of obscured and unobscured growth to the build-up of the supermassive  black holes powering them. 

\section*{Acknowledgements}

The authors would like to thank Jamie Bolton for his detailed and helpful comments on a draft of this manuscript.  We thank Volker Springel for making \textsc{gadget-3} available. We are also grateful to Simon White for granting access to Millennium Simulation data. The plots presented in Figure \ref{densities} use the cube helix colour scheme introduced by \cite{green2011}. This work used the DiRAC Data Analytic system at the University of Cambridge, operated by the University of Cambridge High Performance Computing Service on behalf of the STFC DiRAC HPC Facility (www.dirac.ac.uk). This equipment was funded by BIS National E-infrastructure capital grant (ST/K001590/1), STFC capital grants ST/H008861/1 and ST/H00887X/1, and STFC DiRAC Operations grant ST/K00333X/1. This work also used the DiRAC Data Centric system at Durham University, operated by the Institute for Computational Cosmology on behalf of the STFC DiRAC HPC Facility (www.dirac.ac.uk). This equipment was funded by BIS National E-infrastructure capital grant ST/K00042X/1, STFC capital grants ST/H008519/1 and ST/K00087X/1, STFC DiRAC Operations grant ST/K003267/1 and Durham University.  DiRAC is part of the National E-Infrastructure. LCK acknowledges the support of an Isaac Newton Studentship, the Cambridge Trust and STFC. Support by the  FP7 ERC Advanced Grant Emergence-320596 is gratefully acknowledged.  
\bibliographystyle{mn2e} \bibliography{ref}

\bsp

\appendix

\section{Resolution Tests}
\label{sec:restest}

\begin{figure*}
\includegraphics[trim={0cm 0cm 0cm 21cm},clip,width=2.1\columnwidth]{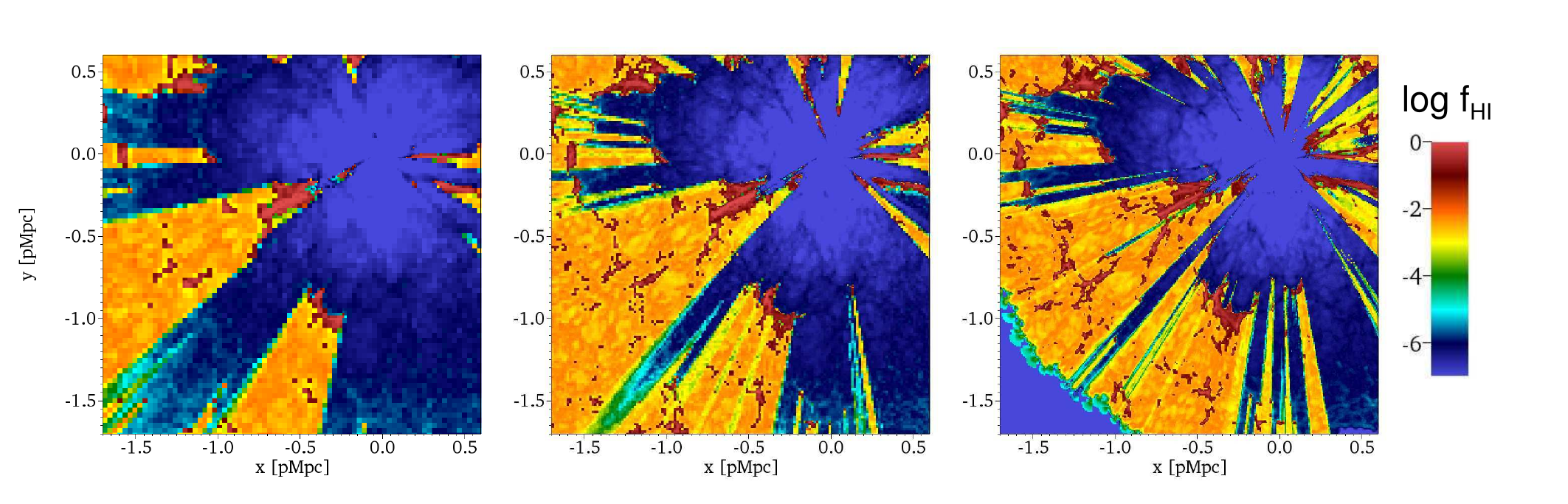}
\caption{Map of the projected neutral hydrogen fraction over a 0.03 pMpc slice in our overdense region at  $t_{\textnormal{\scriptsize{Q}}} = 0.2$ Myr at three different resolutions in a simulation with  $\langle f_{\textnormal{\scriptsize{\ion{H}{i}}}} \rangle_{\textnormal{\scriptsize{init}}} = 0.02$. Left:  $m_{\textnormal{\scriptsize{gas}}} =2.2 \times 10^{7} \, h^{-1} \, M_{\odot}$. Middle:  $m_{\textnormal{\scriptsize{gas}}} =2.8 \times 10^{6} \, h^{-1} \, M_{\odot}$ (default resolution). Right: $m_{\textnormal{\scriptsize{gas}}} =3.5 \times 10^{5} \, h^{-1} \, M_{\odot}$.}
  \label{resfront}
\end{figure*}

\begin{figure*}
\includegraphics[trim={1.5cm 12.5cm 1.5cm 9.5cm},clip,width=2.1\columnwidth]{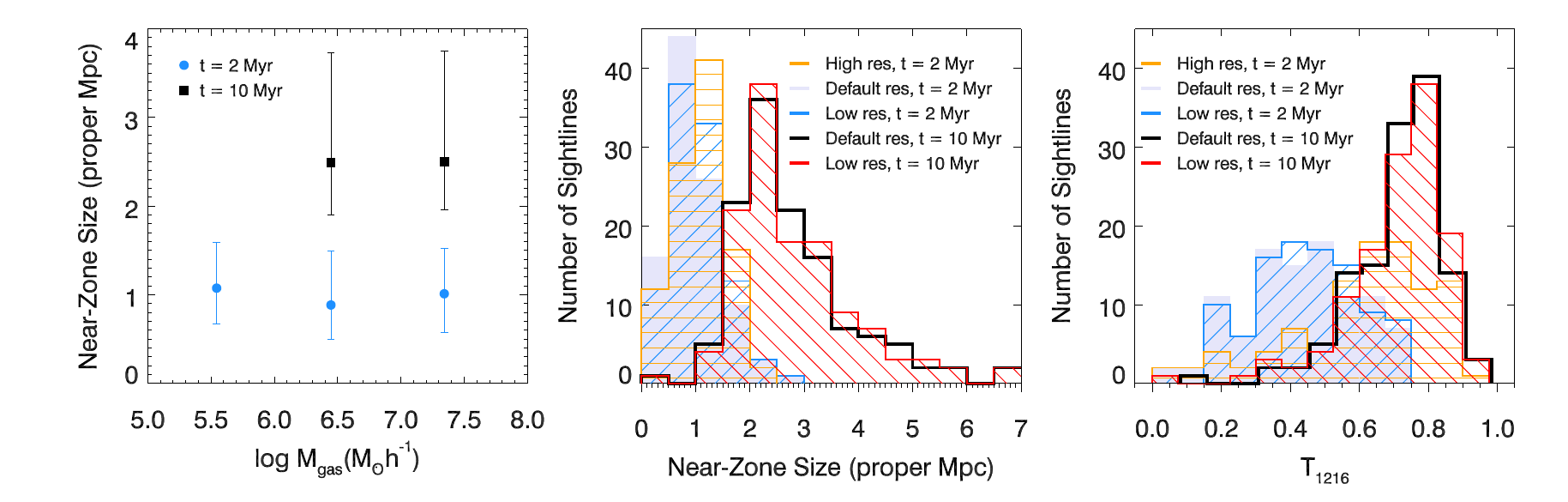}
\caption{Left: median near-zone size with error bars representing the values at the 15$^{\textrm{th}}$/85$^{\textrm{th}}$ percentiles against the mass of the gas particle in the three different resolution simulations we test at $t_{\textnormal{\scriptsize{Q}}} = 2$ Myr and for our default and low resolution simulations at 10 Myr with $\langle f_{\textnormal{\scriptsize{\ion{H}{i}}}} \rangle_{\textnormal{\scriptsize{init}}} = 1.0$. Middle: histogram of the near-zone sizes measured in our overdense region with$\langle f_{\textnormal{\scriptsize{\ion{H}{i}}}} \rangle_{\textnormal{\scriptsize{init}}} = 1.0$ at $t_{\textnormal{\scriptsize{Q}}} = 2$ and 10 Myr. Right: histogram of the transmission at the Ly$\alpha$ rest-frame wavelength measured in our overdense region with $\langle f_{\textnormal{\scriptsize{\ion{H}{i}}}} \rangle_{\textnormal{\scriptsize{init}}} = 1.0$ at $t_{\textnormal{\scriptsize{Q}}} = 2$ Myr and 10 Myr.}
  \label{restest}
\end{figure*}

To study the effect of the numerical resolution of our simulation on the near-zone sizes we measure, we ran two additional hydrodynamic simulations of our overdense region\footnote{In each of these appendices, we perform tests using our overdense region at $z = 7$  which contains the $10^{12.5} M_{\odot}$ halo, unless stated otherwise.}, differing in mass resolution from our fiducial model by a factor of eight, summarised in Table \ref{restable}. For the high resolution run, only a sphere with radius $9 \, h^{-1}$ comoving Mpc was simulated due to computational restraints. The high (low) resolution hydrodynamical simulations were mapped onto correspondingly finer (coarser) grids with respect to our default model. Examples of the resulting ionization fronts in a simulation with $\langle f_{\textnormal{\scriptsize{\ion{H}{i}}}} \rangle_{\textnormal{\scriptsize{init}}} = 0.02$ at  $t_{\textnormal{\scriptsize{Q}}} = 0.2$ Myr  are shown in Figure \ref{resfront}. We find that the scatter in the distance the ionization front has travelled among the different sightlines increases as the resolution is increased.

\begin{table*}
\centering
\begin{tabular}{c|c|c|c}
Resolution&$m_{\textnormal{\scriptsize{gas}}} (\times 10^{6} \, h^{-1}  \, M_{\odot})$&Softening length ($h^{-1}$ kpc)&Size of largest cell ($h^{-1}$ Mpc)\\
\hline
High&0.35&0.625&0.044\\
Default&2.80&1.250&0.089\\
Low&22.0&2.50&0.177\\
\end{tabular}
\caption{Comparison of the three simulations we consider in our resolution tests.}
\label{restable}
\end{table*}

Figure \ref{restest} shows the results of three tests we performed to understand how changing the resolution altered our measurements of the near-zones and the transmission at the Ly$\alpha$ rest frame wavelength. In the left panel, we show the median near-zone size and the values at the  15$^{\textrm{th}}$/85$^{\textrm{th}}$ percentiles as a function of the gas particle mass. The middle panel shows the frequency of different near-zone sizes among the sightlines. We use a snapshot at $t_{\textnormal{\scriptsize{Q}}} = 2$ and 10 Myr in a simulation with  $\langle f_{\textnormal{\scriptsize{\ion{H}{i}}}} \rangle_{\textnormal{\scriptsize{init}}} = 1.0$. For the simulation with  $t_{\textnormal{\scriptsize{Q}}} = 10$ Myr, we only plot the near-zones measured in our default and low resolution simulations, as for this QSO lifetime many of the near-zone sizes are larger than the radius of our highest resolution simulation. We find that, as the resolution is degraded, our simulations predict a similar median and scatter in near-zone sizes. 

We further find that the transmission at Ly$\alpha$ is similar in our default and low resolution simulations at both QSO lifetimes considered here (right panel of  \ref{restest}). The transmission at the Ly$\alpha$ rest-frame wavelength is higher in our highest resolution simulation, most likely due to our star formation prescription. As we increase the resolution of our simulations, we form stars more efficiently and there is  thus  less gas remaining. This will have the largest effect in the densest regions responsible for absorption.  These are the regions that also should be affected by stellar and AGN feedback not included in our default 
simulations.  We note, however, that the median transmission and the scatter in our default simulation is reasonably well matched by a simulation with a more realistic star formation prescription that also includes stellar and AGN feedback (Section \ref{sec:feedback}). 
 
\section{Effect of Feedback}
\label{sec:feedback}

As we expect feedback from stars and AGN to play a role in shaping the environment around galaxies, we ran another simulation of our overdense region which included the effect of  feedback. We have used here the same feedback prescription as in \citet{costa2015}, including supernovae- and AGN-driven outflows, which reproduced well observations of spatially extended cold gas around a $z = 6.4$ QSO. For the supernovae-driven winds, we used a mass loading factor $\eta =1$ and a velocity of 1183 km s$^{-1}$. However, note that our simulation fails to reproduce the fast-moving cold gas found in \citet{costa2015}, perhaps due to differences in the codes used (the moving-mesh code \textsc{arepo} versus the SPH code \textsc{p-gadget3}), a different choice of UV background or the effect of including metal-line cooling. 

\begin{figure*}
  \includegraphics[trim={0cm 0cm 0cm 21cm},clip,width=2.1\columnwidth]{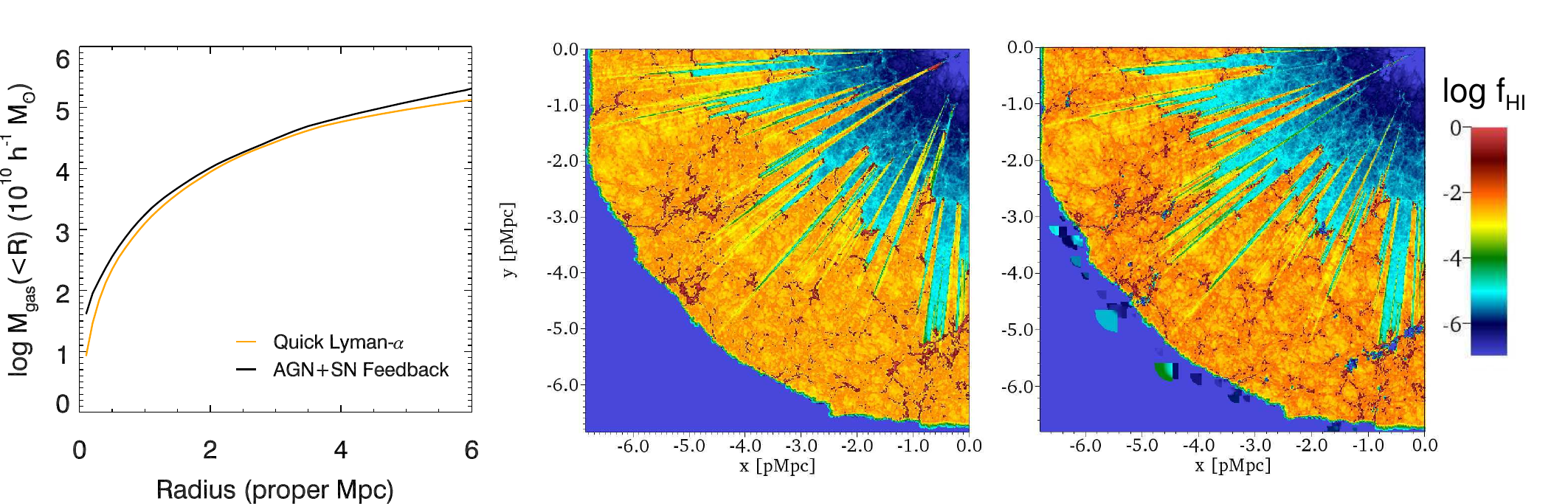}
\caption{Left: total gas mass contained in a sphere with radius specified on the horizontal axis, centred on the host halo for the two simulations. Middle: map of the projected neutral  hydrogen fraction over a 0.03 pMpc slice for our overdense region for the Quick Lyman-$\alpha$ star formation prescription at $t_{\textnormal{\scriptsize{Q}}} = 1$ Myr with background photoionization rate $\log \Gamma_{\textnormal{\scriptsize{\ion{H}{i}}}}$ (s$^{-1}$) = -14.3.  Right:  map of the projected neutral hydrogen fraction for the simulation containing including AGN and SN feedback.}
  \label{fb_fhi}
\end{figure*}

\begin{figure*}
  \includegraphics[trim={1.5cm 12.5cm 1.5cm 9.5cm},width=2.1\columnwidth]{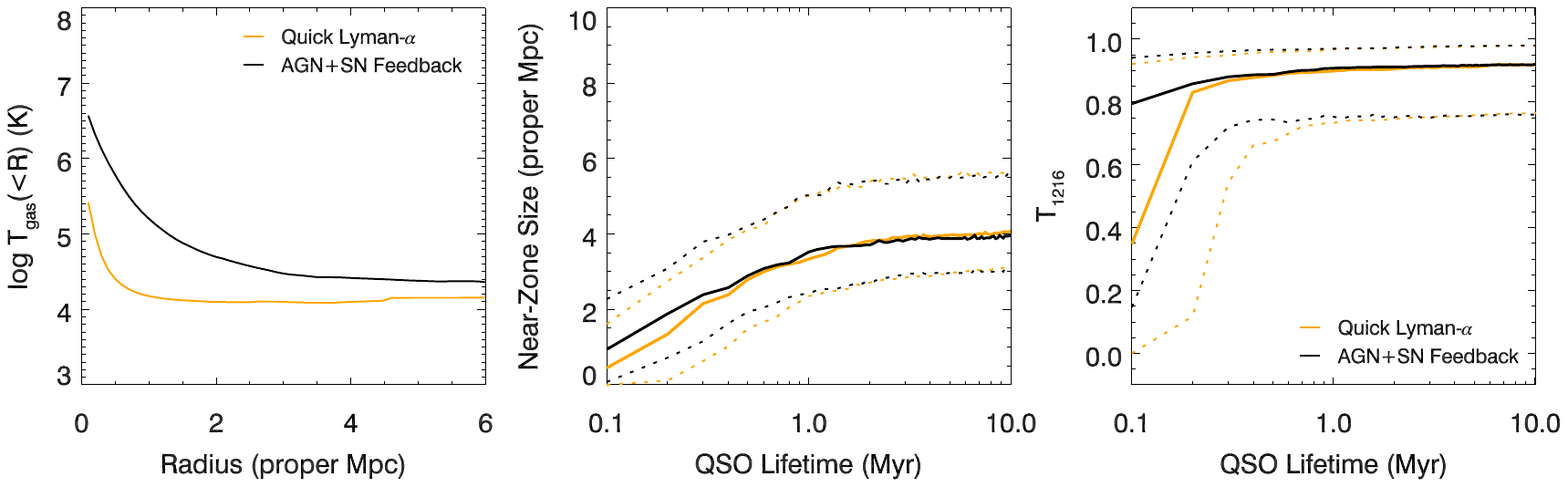}
\caption{Left: mass-weighted average temperature in a sphere with radius specified on the horizontal axis, centred on the host halo for the two simulations. Middle: evolution of the near-zone sizes with time in our overdense region for the models with the Quick Lyman-$\alpha$ star formation prescription and with AGN and SN feedback. The solid line represents the median value and the dashed lines are the values at the  15$^{\textrm{th}}$/85$^{\textrm{th}}$ percentiles. Right: evolution of the transmission at the Ly$\alpha$ rest frame wavelength with time.}
  \label{fb_nz}
\end{figure*}

We find that the shape of the ionization front is remarkably similar to our fiducial model (compare the middle and right panels of Figure \ref{fb_fhi}). There is still a large scatter in the distance that the front has travelled, depending on the sightline. In the middle panel of Figure \ref{fb_nz}, we plot the growth of the median near-zone size with time. We find that, for  $t_{\textnormal{\scriptsize{Q}}} < 0.5$ Myr, the model including feedback shows near-zone sizes with median size $\sim 0.5$ pMpc larger than our fiducial model. This is due to a combination of the galactic winds removing dense gas from the host halo and the thermal feedback resulting in more collisionally ionized gas in the host halo (also visible in some of the larger haloes in the left panel of Figure \ref{fb_fhi}). The right panel of Figure \ref{fb_nz} shows that the temperature of the gas surrounding the host halo is an order of magnitude higher in the case with the AGN and SN feedback compared with the simulation run with Quick Lyman-$\alpha$ star formation.

However, does not have a large effect on the Ly$\alpha$ transmission and for longer QSO lifetimes we find that the median near-zone size and scatter found in our fiducial model is very similar to that in the simulations  with feedback. The star formation prescription we have used in our fiducial model prevents the build-up of high-density gas in the host halo. As well as being computationally efficient, this has the effect of mimicking more physically motivated feedback models which will eject gas from the host halo. We also find evidence for strong damping wings in the simulation including feedback (and note that the \ion{H}{i} photoionization rate used here results in a volume weighted neutral fraction of 0.02 in our fiducial model). 

\section{Ly$\beta$ Near-Zones}

\begin{figure}
\includegraphics[trim={1.5cm 12.5cm 11.5cm 7cm},clip,width=0.95\columnwidth]{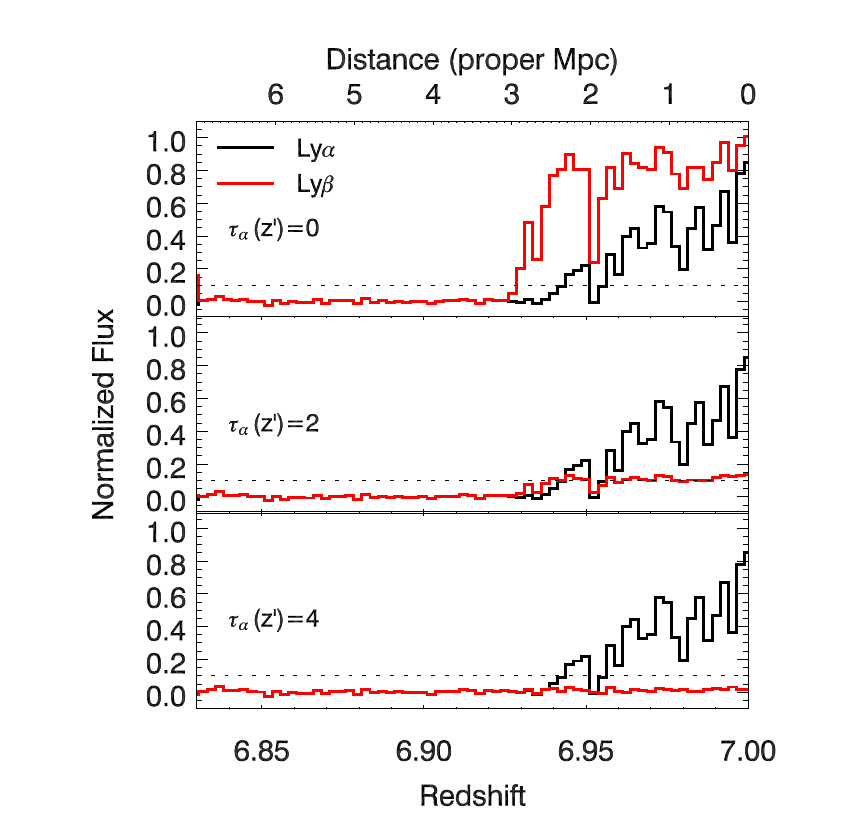}
\caption{Simulated Ly$\alpha$ and Ly$\beta$ spectra along a sightline for initial neutral fraction $\langle f_{\textnormal{\scriptsize{\ion{H}{i}}}} \rangle_{\textnormal{\scriptsize{init}}} = 0.02$ at $t_{\textnormal{\scriptsize{Q}}} = 1$ Myr. In each panel we consider a different value for the optical depth of the Ly$\alpha$ forest at the redshift $z' = [\lambda_{\beta}(1+z)/\lambda_{\alpha}]-1$ corresponding to the Ly$\beta$ absorption. For $\tau_{\alpha}(z') = 0$ (top panel), the contribution from the lower redshift Ly$\alpha$ forest is neglected. For $\tau_{\alpha}(z') = 2,4$ (middle and bottom panels), we experiment with different levels of contribution from the optical depth from the $z=5.75$  Ly$\alpha$ forest. Varying $\tau_{\alpha}(z')$ results in a dramatic change in the observed Ly$\beta$ flux.}
  \label{spectra_lyb}
\end{figure}

We have also investigated the dependence of Ly$\beta$ near-zones on the environment surrounding the host halo. \citet{bolton2007a,bolton2007b} have highlighted  the potential of Ly$\beta$ absorption spectra as a probe of the $z \gtrsim 6$ IGM, as they are expected to take longer to reach  $ R_{\beta}^{\textnormal{\scriptsize{max}}}$ than Ly$\alpha$ near-zones. The corresponding Ly$\alpha$ forest absorption will occur at $z' = [\lambda_{\beta}(1+z)/\lambda_{\alpha}]-1$. For $z = 7$, this corresponds to absorption at $z' = 5.75$. From \citet{becker2015}, the effective optical depth at this redshift will be greater than 2, with some observations returning a lower limit of 7. To test the effect of changing the effective optical depth, we simply increase the Ly$\beta$ optical depth by a constant  $\tau_{\alpha}(z')$ across the whole sightline. Note however taking  $\tau_{\alpha}(z')$ to be constant is an unrealistic assumption as we expect $\tau_{\alpha}(z')$ to vary with changes in the density field, etc. In Figure \ref{spectra_lyb}, we plot the Ly$\alpha$ and Ly$\beta$ spectra along a sightline through our simulation with $\tau_{\alpha}(z') = (0,2,4)$. 

We find that, due to the smaller scattering cross-section of Ly$\beta$, we see more transmitted flux in the Ly$\beta$ spectrum than the Ly$\alpha$ spectrum when $\tau_{\alpha}(z') = 0$. This results in a correspondingly larger near-zone. When we tested the ratio of Ly$\beta$ near-zone sizes to Ly$\alpha$ near-zone sizes, we reproduce the results of \citet{bolton2007b}, where a smaller volume weighted neutral hydrogen fraction gives a larger Ly$\beta$/Ly$\alpha$ near-zone ratio.

However, when we include the additional contribution to the optical depth due to lower redshift  Ly$\alpha$ absorption, we find that the flux  drops accordingly. We still find some  in the Ly$\beta$ region  for $\tau_{\alpha}(z') = 2$, but for $\tau_{\alpha}(z') = 4$ we see no significant transmission of flux along this sightline and  Ly$\beta$ near-zones lose their effectiveness as a probe of the neutral fraction at $z=7$. They may, however, still act a a valuable probe at lower redshift where the ``contamination'' of the Ly$\beta$ region by Ly$\alpha$  absorption is lower. 

\section{Helium Ionization Fronts}

\begin{figure*}
\includegraphics[trim={0cm 0cm 0.5cm 21cm},clip,width=2.1\columnwidth]{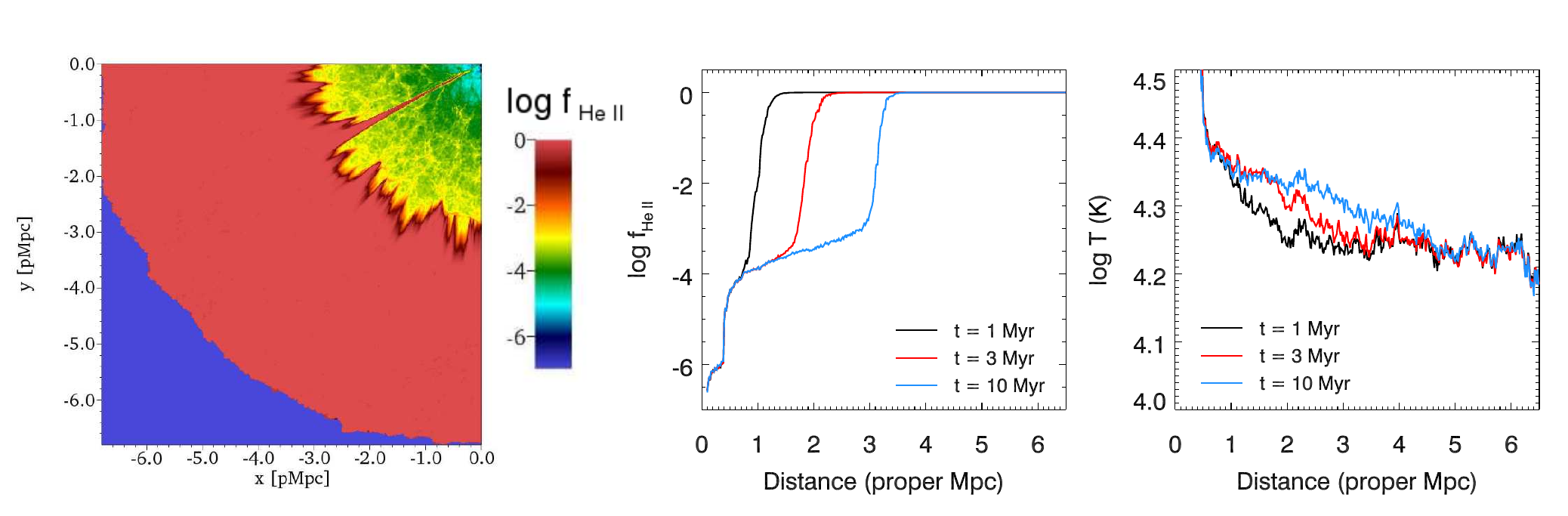}
\caption{Left: projected map of the \ion{He}{ii} fraction after $t_{\textnormal{\scriptsize{Q}}} = 10$ Myr in the simulation with an average initial neutral hydrogen fraction of $\langle f_{\textnormal{\scriptsize{\ion{H}{i}}}} \rangle_{\textnormal{\scriptsize{init}}} = 0.002$.  Middle:  the median neutral \ion{He}{ii} fraction shown at $t_{\textnormal{\scriptsize{Q}}} = 1$ Myr (black), $t_{\textnormal{\scriptsize{Q}}} = 3$ Myr (red) and $t_{\textnormal{\scriptsize{Q}}} = 10$ Myr (blue). Right: the median temperature along our sightlines for the simulation with $\langle f_{\textnormal{\scriptsize{\ion{H}{i}}}} \rangle_{\textnormal{\scriptsize{init}}} = 0.002$ shown at $t_{\textnormal{\scriptsize{Q}}} = 1$ Myr (black), $t_{\textnormal{\scriptsize{Q}}} = 3$ Myr (red) and $t_{\textnormal{\scriptsize{Q}}} = 10$ Myr (blue).}
  \label{heii}
\end{figure*}

We have also investigated how the helium ionization fronts progress and the effect that this as on the temperature of the IGM surrounding the QSO. For a simulation with a large photoionization rate $\Gamma$  (here we present results for $\log \Gamma_{\textnormal{\scriptsize{\ion{H}{i}}}}$ (s$^{-1}$) = -13.7), we find that most of the helium is in \ion{He}{ii}. Otherwise, for smaller $\Gamma$, we see a mixture of \ion{He}{i} and \ion{He}{ii}, with the \ion{He}{i} found in the higher density regions. In the left panel of Figure \ref{heii}, we present a projection of the \ion{He}{ii} fraction after 10 Myr. We find that the front has travelled a much smaller distance than say the front ionizing the \ion{H}{i} (top panel of Figure \ref{fhi_regions}) and the shape is reminiscent of the propagation of a \ion{H}{ii} front into a neutral IGM (bottom panel of Figure \ref{fhi_regions}). 

In the middle panel of Figure \ref{heii}, we show the median \ion{He}{ii} fraction across our sightlines at $t_{\textnormal{\scriptsize{Q}}} = 1, 3$ and 10 Myr. Again, as with the case of the simulations with where the hydrogen in the IGM is initially neutral, we find that the front continues to grow steadily over the 10 Myr we simulate. At $t_{\textnormal{\scriptsize{Q}}} = 1$ Myr, the front has travelled just over 1 pMpc from the host halo. This corresponds to the region of highly ionized hydrogen around the host halo (right panel of Figure \ref{haloes}).  We also find a low \ion{He}{ii} fraction in the immediate vicinity of the host halo where the gas is very hot.

We find that although there is a temperature boost associated with the passage of the \ion{He}{ii} front, it provides an increase of about 20 per cent only (right panel of Figure \ref{heii}). This will not have much of an effect on the recombination rate and hence the near-zone sizes we measure. The radius of the thermal proximity zone is similar in the three regions around our three host haloes and we measure the same temperature at mean density around all three of our host haloes (where $T_{0}$ is the mean volume-weighted gas temperature with a density within 5 per cent of the mean density). At $z=6$, we find that  $\log T_{0} (\textnormal{K}) = 4.30$, within the 1$\sigma$ error bars of the measurement of the temperature around the $z=6$ quasar SDSS J0818+1722 by \citet{bolton2010}.

\label{lastpage}

\end{document}